\documentclass[twocolumn]{aastex631}

\usepackage{csvsimple,longtable,booktabs}
\usepackage{lineno}

\shorttitle{Metallicity Gradients of Star-forming Galaxies}
\shortauthors{Cheng et al.}

\graphicspath{{./}{figures/}}

\begin{document}

\title{Exploring the Gas-Phase Metallicity Gradients of Star-forming Galaxies at Cosmic Noon}

\correspondingauthor{Yingjie Cheng}
\email{yingjiecheng@umass.edu}

\author[0000-0001-8551-071X]{Yingjie Cheng}
\affiliation{University of Massachusetts Amherst \\
710 North Pleasant Street, Amherst, MA 01003-9305, USA}

\author[0000-0002-7831-8751]{Mauro Giavalisco}
\affiliation{University of Massachusetts Amherst \\
710 North Pleasant Street, Amherst, MA 01003-9305, USA}

\author[0000-0002-6386-7299]{Raymond C. Simons}
\affiliation{University of Connecticut \\
196 Auditorium Road, Storrs, CT 06269, USA}

\author[0000-0001-7673-2257]{Zhiyuan Ji}
\affiliation{Steward Observatory, University of Arizona \\
933 N. Cherry Avenue, Tucson, AZ 85721, USA}

\author[0009-0003-9963-6905]{Darren Stroupe}
\affiliation{University of Massachusetts Amherst \\
710 North Pleasant Street, Amherst, MA 01003-9305, USA}

\author[0000-0001-7151-009X]{Nikko J. Cleri}
\affiliation{Department of Physics and Astronomy, Texas A\&M University \\
College Station, TX, 77843-4242 USA}
\affiliation{George P.\ and Cynthia Woods Mitchell Institute for Fundamental Physics and Astronomy, Texas A\&M University \\
College Station, TX, 77843-4242 USA}

\begin{abstract}

We explore the relationships between the [O/H] gas-phase metallicity radial gradients and multiple galaxy properties for 238 star-forming galaxies at 0.6$<$z$<$2.6 selected from the CANDELS Ly$\alpha$ Emission at Reionization (CLEAR) survey with stellar mass 8.5 $<\log~M_*/M_{\odot}<$ 10.5. The gradients cover the range from -0.11 to 0.22 dex $\mathrm{kpc}^{-1}$, with the median value close to zero. We reconstruct the non-parametric star-formation histories (SFHs) of the galaxies with spectral energy distribution modeling using $Prospector$ with more than 40 photometric bands from {\it HST}, Spitzer and ground-based facilities. In general, we find weak or no correlations between the metallicity gradients and most galaxy properties, including the mass-weighted age, recent star formation rate, dust attenuation, and morphology as quantified by both parametric and non-parametric diagnostics. We find a significant but moderate correlation between the gradients and the `evolutionary time', a temporal metric that characterizes the evolutionary status of a galaxy, with flatter gradients observed in more evolved galaxies. Also, there is evidence that galaxies with multiple star-formation episodes in their SFHs tend to develop more negative gas-phase metallicity gradients (higher [O/H] at the center). We conclude that gas kinematics, e.g. radial inflows and outflows, is likely an important process in setting the gas-phase metallicity gradients, in addition to the evolution of the SFH radial profile. Since the gradients are largely independent on the galaxies' physical properties, and only weakly dependent on their SFH, it would appear that the time-scale of the gas kinematics is significantly shorter than the evolution of star formation.

\end{abstract}

\keywords{galaxies: high-redshift --- ISM: abundances --- galaxies: evolution}

\section{Introduction} \label{sec:intro}

Galaxies form and evolve in a gas ecosystem, with metals in the gas phase continuously produced, transported and consumed throughout their evolution. Star formation, supernovae and neutron star mergers \citep[e.g,][]{2017ARNPS..67..253T} contribute to the local metallicity enrichment of the interstellar medium (ISM). Meanwhile, the stellar remnants and the evolution of low-mass stars lead to the consumption of metals \citep[e.g,][]{2006ApJ...653.1145K, 2016ApJ...821...38S}. The distribution and evolution of gas-phase metallicity gradients provide important insights into the star formation history (SFH) of galaxies as well as how they evolve over time. 

If a galaxy were a closed box without any gas exchange, and thus metal exchange with the external environment, one would expect that the gas-phase metallicity gradient would tend to follow the radial distribution of past generations of stars as they formed. However, metals can be lost through galactic winds and outflows \citep[e.g,][]{2018Galax...6..138R, 2019ApJ...886...74M, 2022ApJ...930..146W}, accreted from the surrounding intergalactic medium (IGM) and/or the CGM \citep[e.g,][]{2011ApJ...731...11C, 2017MNRAS.468.4170M, 2021MNRAS.505.4655S, 2022ApJ...934..100S}, and re-distributed either through internal dynamical mechanisms or through interactions with other galaxies such as mergers \citep[e.g,][]{2021ApJ...907..110B, 2021MNRAS.505..339M}. All these factors can effectively change the distribution and evolution of metals in galaxies, and thus add to the complexity of interpreting the shape of the gas-phase metallicity gradients and their relationship with the galaxies' stellar mass and star-formation rate (SFR) gradients.

Gas-phase metallicity gradients have been extensively studied through observations in the local universe since the 1970s \citep{1971ApJ...168..327S, 1983MNRAS.204...53S}. In the local universe, early works showed that the metallicity decreases approximately exponentially with the galactocentric radius (namely showing a negative gradient) in disk-like galaxies (e.g. \citealt{1992ApJ...390L..73Z, 1995ApJ...438..170H}). With the development of contemporary instruments and large integral-field spectroscopic surveys --- e.g. CALIFA \citep{2012A&A...538A...8S} and MaNGA \citep{2015ApJ...798....7B} ---  the measurements of gas-phase metallicity gradients have been dramatically expanded. Local galaxies are consistently reported to have negative radial gradients in gas-phase metallicity, with typical values ranging from 0 to -0.1 dex $\mathrm{kpc}^{-1}$ \citep[e.g,][]{2019ARA&A..57..511K, 2019ApJ...887...80K, 2020ARA&A..58...99S}. Flat or positive gradients are also found occasionally, typically for low-mass and late-type galaxies where the estimation of age and metallicity is highly degenerated, or for galaxies in dense environments \citep[e.g,][]{2019MNRAS.489.1436L, 2020ARA&A..58...99S}. 
By measuring the gas fraction and the local escape velocity, \citet{2018ApJ...852...74B} were able to reproduce the observed radial metallicity gradients for a large sample of local disk galaxies. This result indicated that the gas-phase metallicity is a consequence of local star formation history, and momentum-driven outflows help shape the internal metallicity of star-forming galaxies.

Measuring metallicity gradients in high-z galaxies is challenging because galaxies were smaller in the past and the available sensitivity and angular resolution are still limited compared to observations in the local universe. To quantify the systematics encountered by observing poorly-resolved galaxies, \citet{2020MNRAS.495.3819A} used synthetic observations of numerical galaxy simulations to demonstrate that the observed metallicity gradient is artificially flattened at low spatial resolution and low signal-to-noise ratio. For this reason, the sizes of the samples at high redshift with robust measurements of metallicity gradients have traditionally remained small. Thanks to the sensitive and high-resolution near-infrared spectrographs onboard the {\it Hubble Space Telescope} and, more recently, with JWST, as well as many large ground-based facilities, the observations of gas-phase metallicity gradients in high-z galaxies have been expanding in the last decade \citep{2016ApJ...827...74W, 2017ApJ...837...89W, 2019ApJ...882...94W, 2020MNRAS.492..821C, 2020ApJ...900..183W, 2021MNRAS.500.4229G, 2021ApJ...923..203S, 2022ApJ...938L..16W}. Besides the metallicity gradient itself, these studies have also investigated the relationship between the metallicity gradient and galaxy properties, including the stellar mass, size, and morphology. 

Gas-phase metallicity gradients have been reported to correlate with the stellar mass especially in the local universe, but how exactly the two properties are correlated with each other remains inconclusive. Some suggested that high-mass (typically $\log(M/M_{\odot})\gtrsim10.5$) galaxies tend to have slightly more negative metallicity gradients compared with low-mass galaxies (e.g. \citealt{2017MNRAS.469..151B, 2018MNRAS.478.4293C, 2018MNRAS.479.5235P}).
In contrast, \citet{2020A&A...636A..42M} reported a slight steepening in metallicity gradient with stellar mass at $log(M/M_{\odot})$ = 9$\sim$10.25, and then a flattening towards $log(M/M_{\odot})$ = 11. \citet{2021MNRAS.504...53S} and \citet{2022MNRAS.512.3480G} reached similar conclusion at z$\sim$0 and z$\sim$1.5 respectively. An analysis of the \textit{Illustris TNG} high-resolution cosmological simulations also suggested  that at z$<$2, galaxies with larger stellar mass generally have flatter (i.e. less negative) gas-phase metallicity gradients \citep{2021MNRAS.506.3024H}.
\citet{2021ApJ...923..203S} reported that massive galaxies have flatter gradients and lower mass galaxies have more positive gradients. This dependence may indicate the redistribution of metals in the early stages of galaxy formation, such as powerful feedback effects, radial flows, and stochastic accretion from minor mergers in the early universe \citep{2019A&ARv..27....3M}.

The dependence of gas-phase metallicity gradients on galaxy sizes is also disputed. Some works reported  that larger galaxies tend to have steeper (more negative) metallicity gradients \citep{2018MNRAS.478.4293C, 2020MNRAS.491.3672B}, while some others reported no dependence on size of  gas metallicity gradients for their samples (e.g. \citealt{2018A&A...609A.119S, 2021ApJ...923..203S}).

Because the structure of galaxies depends on the stellar mass, it is natural to test the connection between morphology and metallicity gradients. Such a test has been carried out in the local universe. For example, \citet{2016A&A...587A..70S} reported that earlier type spirals show flatter gradients than the later type ones. How this dependence is affected by the redshift evolution of morphology, however,  remains poorly understood.

Gas-phase metallicity gradients also provide crucial constraints on theoretical models of galaxy formation and evolution. For instance, the observed negative radial gradients in both the gas-phase and stellar  metallicities at low redshift has been interpreted as strong evidence for the scenario of inside-out galaxy formation. In this model, the center of the galaxy forms first, undergoes more generations of star formation, and thus accumulates more metals compared with the outskirts. Although the negative trend is qualitatively easy to understand, quantitatively reproducing the observed gradients as well as their dependence on galaxy properties requires detailed modeling that includes assumptions on the dependence of star formation activity on the local gas distribution (e.g. the star formation law or Kennicutt-Schmidt Law, \citealt{1998ApJ...498..541K}) as well as the SFR radial dependence \citep[e.g,][]{2019MNRAS.487..456B, 2021MNRAS.502.1967K, 2021MNRAS.502.5935S}. 

Recent theoretical efforts to understand and reproduce the observed metallicity gradients in galaxies have relied mostly on semi-analytic models (e.g., \citealt{2019MNRAS.487.3581F, 2021MNRAS.503.4474Y}) and cosmological hydrodynamical simulations (e.g., \citealt{2019MNRAS.482.2208T, 2021MNRAS.506.3024H, 2021MNRAS.505.4586B}). Many works have succeeded in reproducing the present-day metallicity gradient in our Milky Way (e.g. \citealt{2014AJ....147..116H, 2018MNRAS.481.1645M, 2021A&A...656A.156Q}), but analyzing the metallicity in high-z galaxies with differing evolutionary histories is far more complicated. The results can be affected by which processes are being considered or emphasized, but a general conclusion of these works is that there is only a weak steepening trend in metallicity gradients as galaxies evolve towards the present day from z $\leq$ 4 \citep{2020MNRAS.492..821C}.  \citet{2021MNRAS.502.5935S} presented a theory of the evolution of gas-phase metallicity gradients in galaxies, derived from first principles, which is based on the unified galactic disc model given by \citealt{2018MNRAS.477.2716K}. This theory is able to explain the dependence of metallicity gradients on the stellar mass and kinematics of galaxies and  reproduces the observations in the local Universe.

Recently, \citet{2021ApJ...923..203S} measured gas-phase metallicity gradients for a large sample of over 200 star forming galaxies at $0.6<z<2.6$. The data were obtained from \textit{HST}/WFC3 G102 slitless grism spectra as part of the Cycle 23 CANDELS Lyman $\alpha$ Emission at Reionization survey (CLEAR, \citealt{2023arXiv230309570S}).
The authors find a broad dispersion of values of the gradients, ranging from -0.11 to 0.22 dex $\mathrm{kpc}^{-1}$ with essentially no apparent trend or correlation (Pearson correlation coefficient $|\rho |<0.2$) with galaxy properties such as stellar mass, SFR and morphology.

In this paper, we follow-up on the work by \citet{2021ApJ...923..203S} and expand the investigation to explore the dependence of the gradients with the star-formation history of the galaxies, which we have derived in non-parametric form using the software package $Prospector$ \citep{2021ApJS..254...22J}. The idea behind this investigation is that the presence or absence of such relationships will help provide a better physical understanding on the formation and assembly history of high-z galaxies. 

The rest of the paper is organized as follows. Section \ref{sec:data} describes our data and sample selection. Section \ref{sec:fit} shows the spectral energy distribution fitting with $Prospector$. Section \ref{sec:sfh}, \ref{sec:color} and \ref{sec:morph} investigate the relationship between the gas-phase metallicity gradients and multiple galaxy properties, including the SFH, colors and the morphology. Section \ref{sec:dis} discusses the implications of this study in terms of galaxy evolution, and Section \ref{sec:sum} provides a brief summary of our main conclusions. 

Throughout this paper, we adopt a flat $\Lambda$CDM cosmology model with $H_0=70~km~s^{-1}Mpc^{-1}$ and $\Omega_{m0}=0.3$.

\section{Data and Sample Selection} \label{sec:data}
\subsection{Gas-phase metallicity gradients}

\begin{figure}
\plotone{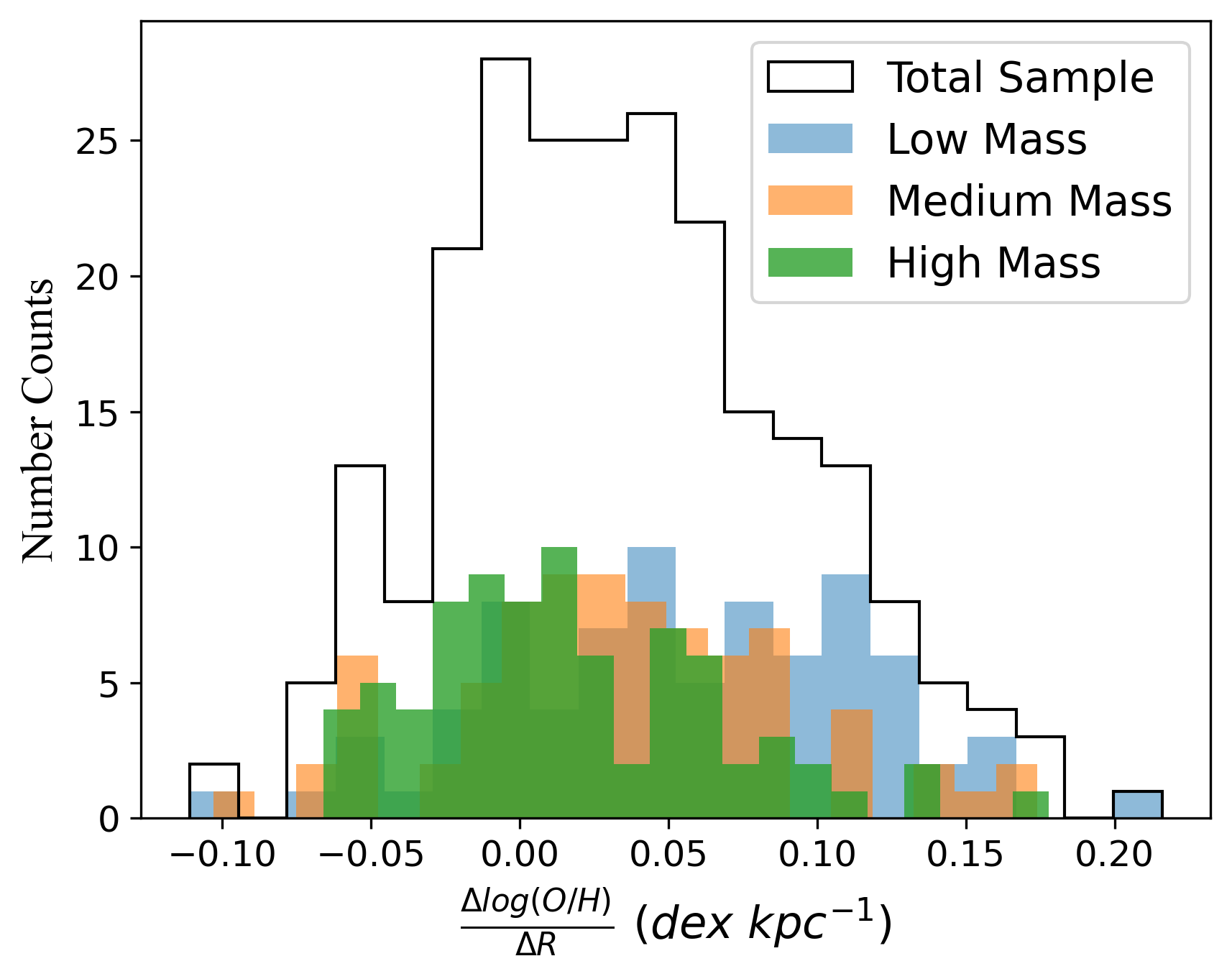}
\caption{The histograms of the gas-phase metallicity gradients taken from \citealt{2021ApJ...923..203S} for the whole sample as well as for three stellar mass bins of equal sample size (8.5$<\log~M_*/M_{\odot}<$9.5 for the low mass bin, 9.5$<\log~M_*/M_{\odot}<$9.9 for the medium mass bin, and 9.9$<\log~M_*/M_{\odot}<$10.5 for the high mass bin). There is evidence that galaxies with higher stellar mass tend to grow slightly more negative metallicity gradients. \label{fig:hist}}
\end{figure}

In this paper we use the measures of gas-phase metallicity gradients of high-z star-forming galaxies in the GOODS-S and GOODS-N fields made by \citet{2021ApJ...923..203S} and expand the analysis of their correlation with the star formation history of the galaxies. A full description of the sample selection, observations, data reduction and analysis, as well as the creation of the line emission maps and measure of the metallicity gradients can be founds in Section 2 of \citet{2021ApJ...923..203S}. 
The sample and the data of this work are exactly the same as those in that paper. To illustrate the data, in Appendix \ref{appendix:a} we provide the emission line maps of one galaxy as an example, color coded by pixel SNR, together with extracted 1D spectra from selected extraction boxes. \citet{2021ApJ...923..203S} provide a complete description of the line fitting algorithms and of the azimuthal averaging procedures adopted to measure the metallicity gradients.

The data were obtained from deep near-infrared Hubble Space Telescope slitless spectroscopy as a part of the CLEAR survey (see \citealt{2023arXiv230309570S}). The footprint of this survey covers 12 pointings in the CANDELS GOODS-N and GOODS-S fields \citep{2011ApJS..197...35G,2011ApJS..197...36K}, providing deep HST/WFC3 G102 grism spectroscopy and companion HST/WFC3 F105W direct imaging. All publicly available HST/WFC3 G102 and G141 grism observations that overlap with the CLEAR footprint are also included to extend the spectral coverage and maximize the depth. The redshift fitting and grism spectral analysis are conducted with the software $Grizli$ \citep{2019ascl.soft05001B}, and the emission line fluxes are derived from Flexible Stellar Population Synthesis (FSPS) templates \citep{2009ApJ...699..486C}. Emission line maps are then created by drizzling the contamination- and continuum-subtracted 2D spectral beams.

The sample of galaxies were selected based on high S/N line ratio criteria ($\geq5\sigma$ integrated detection in at least 2 of the strong lines in $[\mathrm{OIII}]~\lambda\lambda 4958, 5007,~[\mathrm{OII}]~\lambda 3727$, and $\mathrm{H} \beta$) and for having a radial extent that allowed spatially resolved emission line maps. The metallicity gradients were derived from line ratio diagnostics of these maps. Galaxies with detected X-ray emission consistent with a Active Galactic Nuclei (AGN) have been removed from the sample. The total sample contains 238 star-forming galaxies at 0.6 $<z<$ 2.6 across a mass range of 8.5 $<~\log~M_*/M_{\odot}~<$ 10.5. Around 96\% of the sample are consistent with a flat or positive metallicity gradient within $3\sigma$ confidence while only 4\% show a negative gradient. The number distribution of the derived metallicity gradients is shown in Figure \ref{fig:hist}. We show the distribution both for the total sample, and for three equally divided sub-samples according to the stellar mass.

\subsection{Photometry and morphology data}

We have modeled the spectral energy distributions (SEDs) of the galaxies in our sample and reconstructed their SFH in non-parametric form with the package $Prospector$ using the extended CANDELS (Cosmic Assembly Near-infrared Deep Extragalactic Legacy Survey, \citealt{2011ApJS..197...35G}) photometric catalogs in the GOODS fields \citep{2004ApJ...600L..93G}. These include over 40 photometric bands from both ground-based and space-borne instruments, ranging from the rest-frame ultraviolet (UV) portion of the spectrum to the mid-infrared (MIR) one. For the GOODS-N field, we use the photometric catalog from \citet{2019ApJS..243...22B}. This  includes broadband data from the observed UV (U band from KPNO and LBC), optical (HST/ACS F435W, F606W, F775W, F814W, and F850LP), NIR (HST/WFC3 F105W, F125W, F140W, and F160W; Subaru/MOIRCS Ks; CFHT/Megacam K; NIR and MIR Spitzer/IRAC 3.6, 4.5, 5.8, and 8.0 $\mu$m), plus 25 consecutive medium bands from the SHARDS survey at $\lambda$ = 500–950 nm \citep{2019ApJS..243...22B}. This yields a total of 42 photometric bands used in the $Prospector$ SED modeling.

For the GOODS-S field we adopt the ASTRODEEP-GS43 photometric catalog made by \citet{2021A&A...649A..22M}, which is built and improved upon the previously released CANDELS catalog. To add to the CANDELS photometry in 7 bands (CTIO U, Hubble Space Telescope WFC3 and ISAAC-K), they also obtained measures in the other 36 bands (VIMOS, HST ACS, HAWK-I Ks, Spitzer IRAC, and 23 from Subaru SuprimeCAM and Magellan-Baade Fourstar), all the aperture-matched with the template-fitting package TPHOT \citep{2021A&A...649A..22M}.
This yields a total of 43 wavelength bands (25 wide and 18 medium filters).

Finally, \citet{2015ApJS..221....8H} presented a catalog of visual-like H-band morphologies of CANDELS galaxies estimated from Convolutional Neural Networks. We use their morphology classifications along with the best-fit H-band S\'ersic indexes provided by the CANDELS catalog \citep{2012ApJS..203...24V} to explore the dependence of metallicity gradients on the morphological properties of galaxies in Section \ref{sec:morph}.

\section{SED fitting with Prospector} \label{sec:fit}
With high-quality, multi-band photometry data, we use $Prospector$ \citep{2021ApJS..254...22J} to perform SED modeling and  reconstruct the SFHs of selected galaxies.
$Prospector$ is a Python-based package to conduct principled Bayesian inference of stellar-population properties by fitting photometric and/or spectroscopic data to spectral population synthesis models. The code enables one to obtain the galaxies' SFH in a non-parametric form, as opposed to a pre-assumed analytical functional form, which has been shown to minimize systematics in the measures of stellar mass, SFR and stellar age  \citep{2019ApJ...879..116I, 2019ApJ...876....3L, 2021arXiv211004314L, 2021essp.confE..59T, 2022ApJ...927..199D}. The fitting is based on the FSPS code \citep{2009ApJ...699..486C}, which uses the MIST stellar isochrone library \citep{2016ApJ...823..102C} and the MILES spectral library \citep{2011A&A...532A..95F}. Also, the code self-consistently models nebular emission (line plus continuum) using a photo-ionization model described in \citealt{2017ApJ...840...44B}. 

Specifically, we adopt the non-parametric SFH model which consists with a certain number of bins in look-back time from the time of observations where the past SFR is estimated, using a prior to control the smoothness of the transition of the SFR from one bin to the next. Following other authors (e.g. \citealt{2021ApJS..254...22J,2022ApJ...926..134T, 2022ApJ...935..120J}), who have extensively tested the procedure, we also adopted the continuity prior described in \citealt{2019ApJ...876....3L} in our analysis to emphasize smoothness in the star formation rate (SFR). The code directly fits for $\Delta~log(SFR)$ between adjacent time bins and explicitly weights against sharp transitions (see \citealt{2019ApJ...876....3L} for details). We have also experimented with the Dirichelet prior, however, finding that the choice of the prior only minimally affects our results and conclusions.

We adopt 9 lookback time bins in this study, with the SFR being constant within each bin. The time bins are adjusted based on the age of the universe at the given redshift for each galaxy. The first lookback time bin is fixed at $0<t<30\ \mathrm{Myr}$ to capture the recent episodes of star formations, and the last bin is set to be $85\%-100\%$ of the age of universe at any given redshift. All the intervening bins are evenly spaced in $log(t)$ scale. To test the possible dependence of the SED fitting results on the time binning procedure, we reran the fitting for all the GOODS-N galaxies in our sample with 7 look-back time bins and compared with the 9-bin results (see Appendix \ref{appendix:b} for details). After comparing the recovered SFH plots and galaxy physical properties in the two cases, we concluded that the fitting results with 7 bins and 9 bins are fully consistent with each other. This analysis is also in agreement with previous work reporting that the fitting results are generally insensitive to the number of time bins when there are more than 5 bins \citep{2019ApJ...876....3L}. Following other groups \citep[e.g,][]{2022ApJ...926..134T,2022ApJ...935..120J} for the rest of the paper we will be presenting our results obtained using the 9-bin SFH. 
The detailed parameter settings are described as follows.

\subsection{Other adopted assumptions} \label{subsec:set}

\begin{figure*}
\plotone{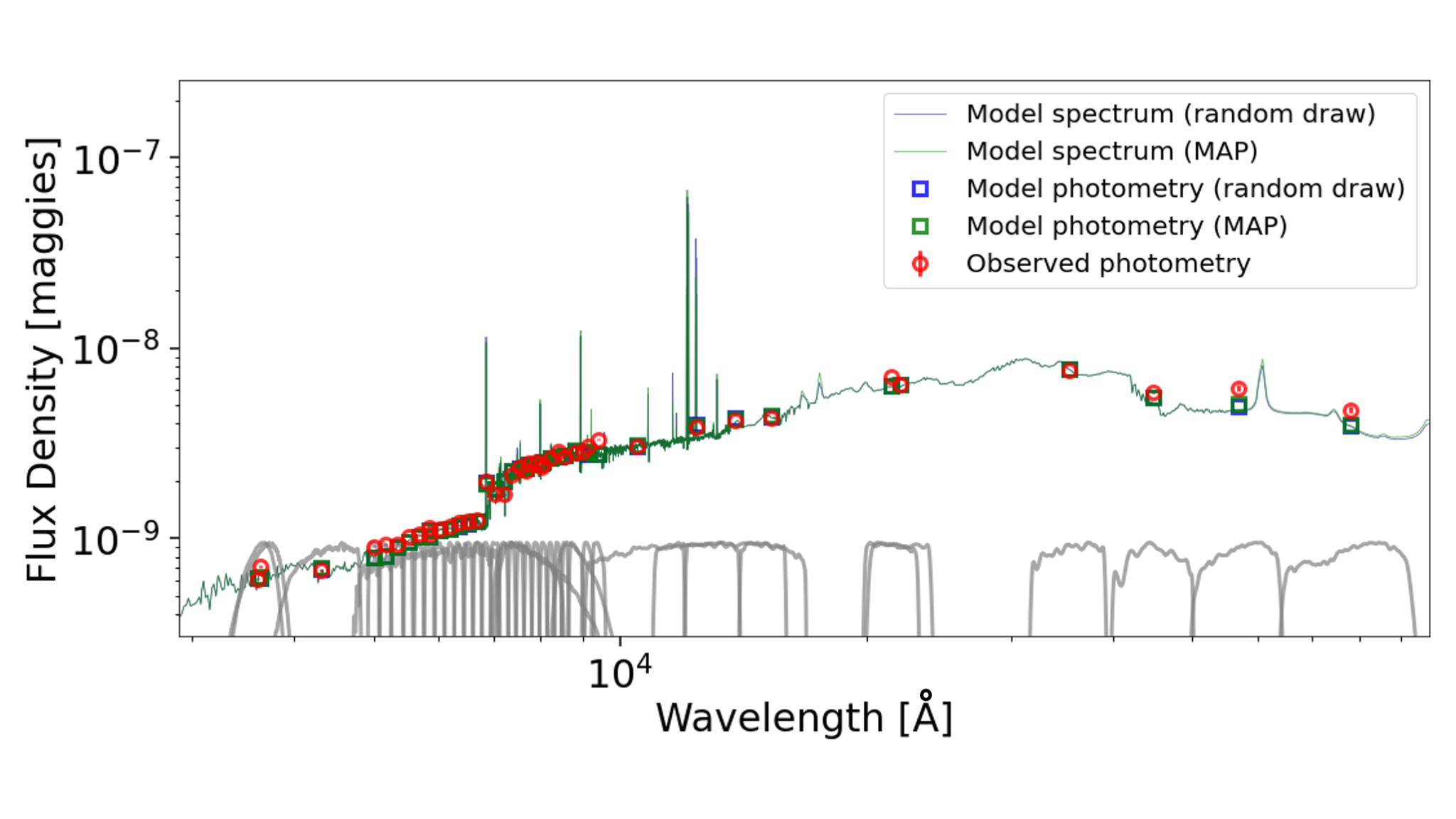}
\caption{An example of the best-fit SED for source No.16607 in the GOODSN field. The wavelengths are in units of $\mathring{A}$, the flux densities are in units of maggies (Jy/3631). The red circles show the observed data, and the gray lines show the transmission curves of the filters. A random model drawn from the sampling chain is plotted in blue, while the model at the maximum posterior probability (MAP) is plotted in green.
\label{fig:spec}}
\end{figure*}

In this work we adopt the Kroupa's initial mass function \citep{2001MNRAS.322..231K} and the Calzetti's dust attenuation law \citep{2000ApJ...533..682C}. The diffuse dust V-band optical depth (the `dust2' parameter in FSPS) is fitted with an uniform prior between 0 and 4. We also fix the redshift at the best-fit grism redshift given by \citet{2021ApJ...923..203S}. The stellar metallicity (the `logzsol' parameter in FSPS) is left as a free parameter, but with a strong Gaussian prior centered at the average observed gas-phase metallicity, i.e. obtained by averaging the observed gradient within 2$R_e$, and width equal to the corresponding uncertainty.

The sampling is performed with the dynamical nested sampling code $Dynesty$ \citep{2020MNRAS.493.3132S}, which allows us to control the effective time resolution and to sample preferentially near the given posterior. Though costly in terms of time, the $Dynesty$ sampling method has been proved to be more accurate than the traditional MCMC sampling method, especially when a non-parametric SFH is assumed \citep{2019ApJ...876....3L}.

\subsection{Results} \label{subsec:res}

\begin{figure}
\plotone{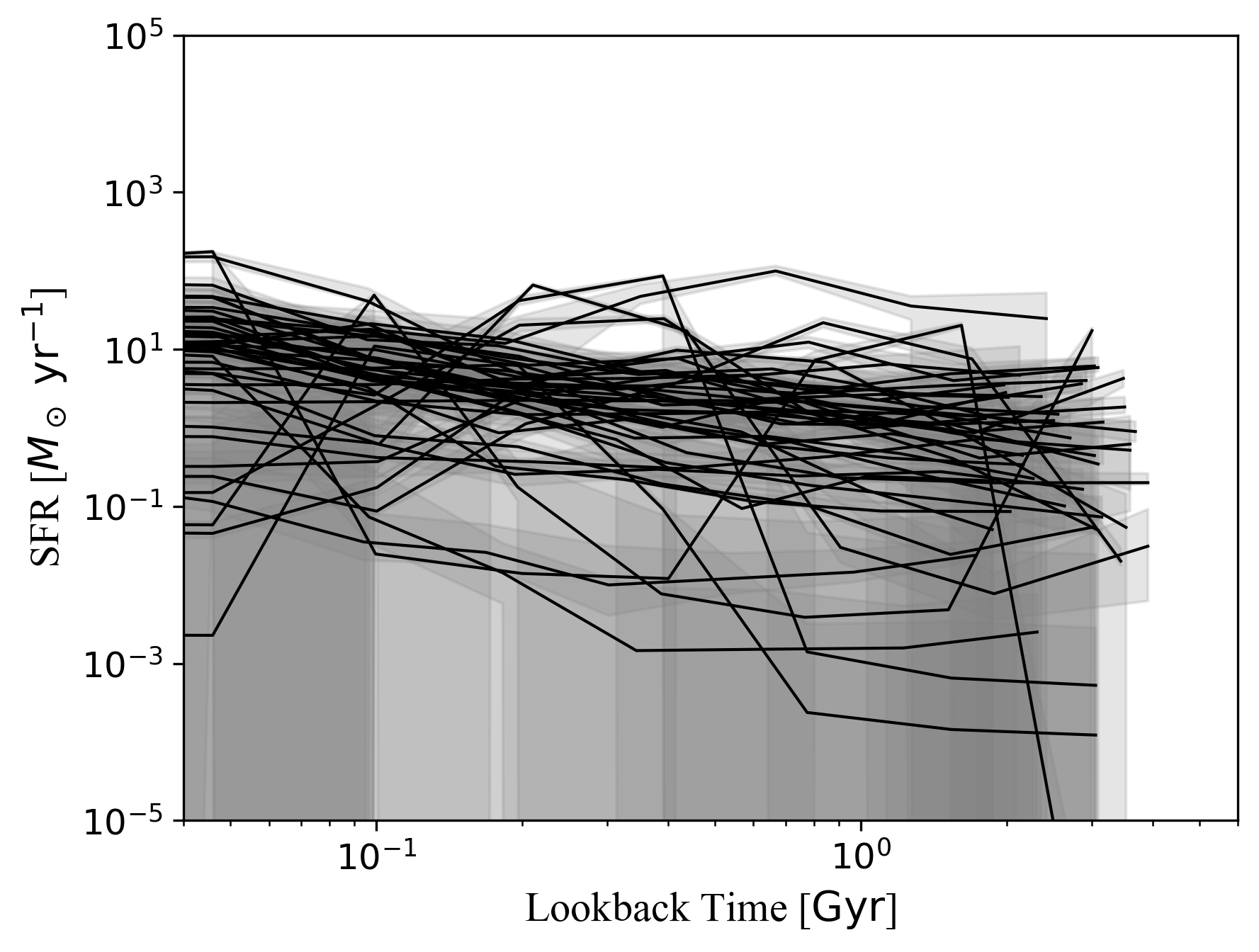}
\caption{The stacked star formation history plots with 9 lookback time bins \label{fig:stack}}
\end{figure}

We do not directly test for the robustness of $Prospector$'s SFH reconstruction here, because such tests have been done in detail by several different teams of investigators, including the builders of the code themselves (e.g, \citealt{2017ApJ...837..170L, 2021ApJS..254...22J, 2022ApJ...927..170T, 2022ApJ...935..120J} and reference therein). The general result from these testing procedures, which rely on synthetic galaxies from cosmological hydrodynamics simulations, such as IllustrisTNG \citep{2018MNRAS.473.4077P}, is that $Prospector$ generally yields accurate and relatively precise SFH. 

In this paper, we have tested the stability of $Prospector$ against variations in the quality and number of the input photometric bands to take into account the fact that not all galaxies have the same photometric coverage and that SNRs in the various bands vary from galaxy to galaxy. First, we test for the effect of missing bands by rerunning the fitting with less photometric bands. In comparison with the $\sim$40 bands adopted before, we use a set of $\sim$20 photometric bands provided by the 3DHST survey \citep{2016ApJS..225...27M}, including data from the HHDFN U, B, V, R bands; HST/ACS F435W, F606W, F775W, and F850LP; HST/WFC3 F125W, F140W, and F160W; MODS J, H, Ks; and Spitzer/IRAC 3.6, 4.5, 5.8, and 8.0 $\mu$m bands. Although the results are highly consistent for most galaxies in our sample, a few of them show different features in the best-fit SFH (See Appendix \ref{appendix:b}). 
Therefore, to secure high-quality SFH reconstructions  we have visually inspected the photometry of our sample to only select galaxies with high-quality, densely sampled photometry. Specifically, we have removed galaxies that have more than 5 missing bands and/or have large photometric error bars, i.e. SNR$\approx 5$ or less, in ground-based observations. This vetting procedure leaves us with a sample of 119 star-forming galaxies, about half of the original sample size.

\begin{figure*}
\plotone{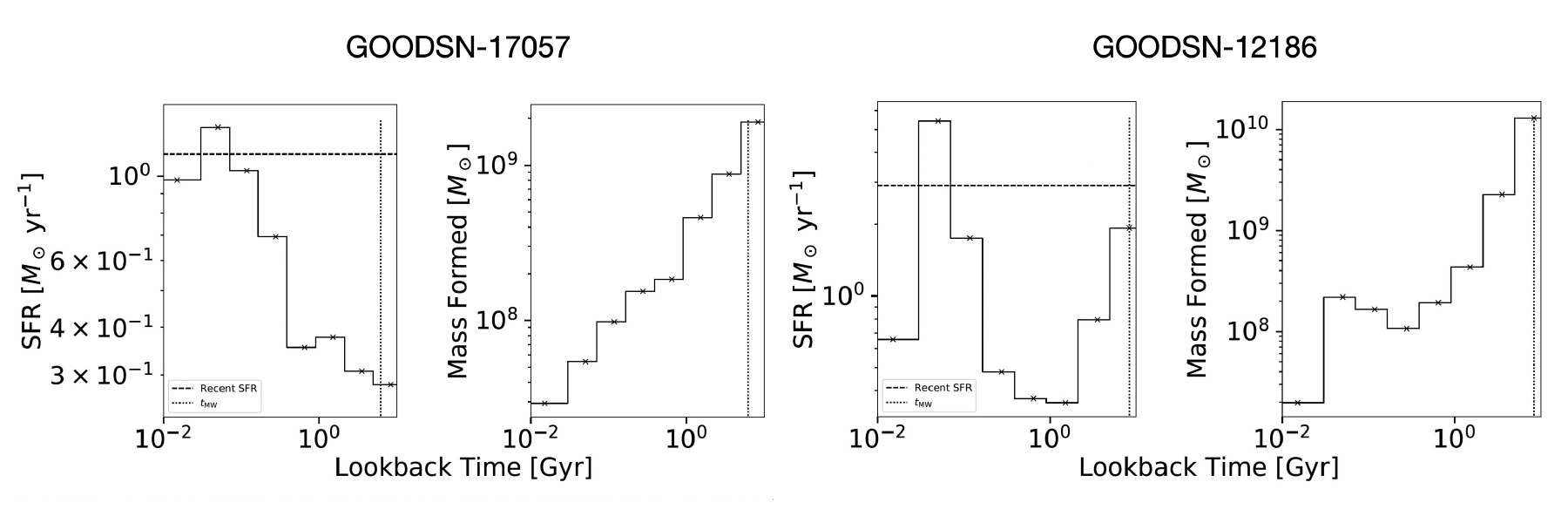}
\caption{Example plots for galaxies with single (left) or multiple (right) star formation episodes in their SFHs. 
\label{fig:ex}}
\vspace{1cm}
\end{figure*}

\begin{figure*}
\plottwo{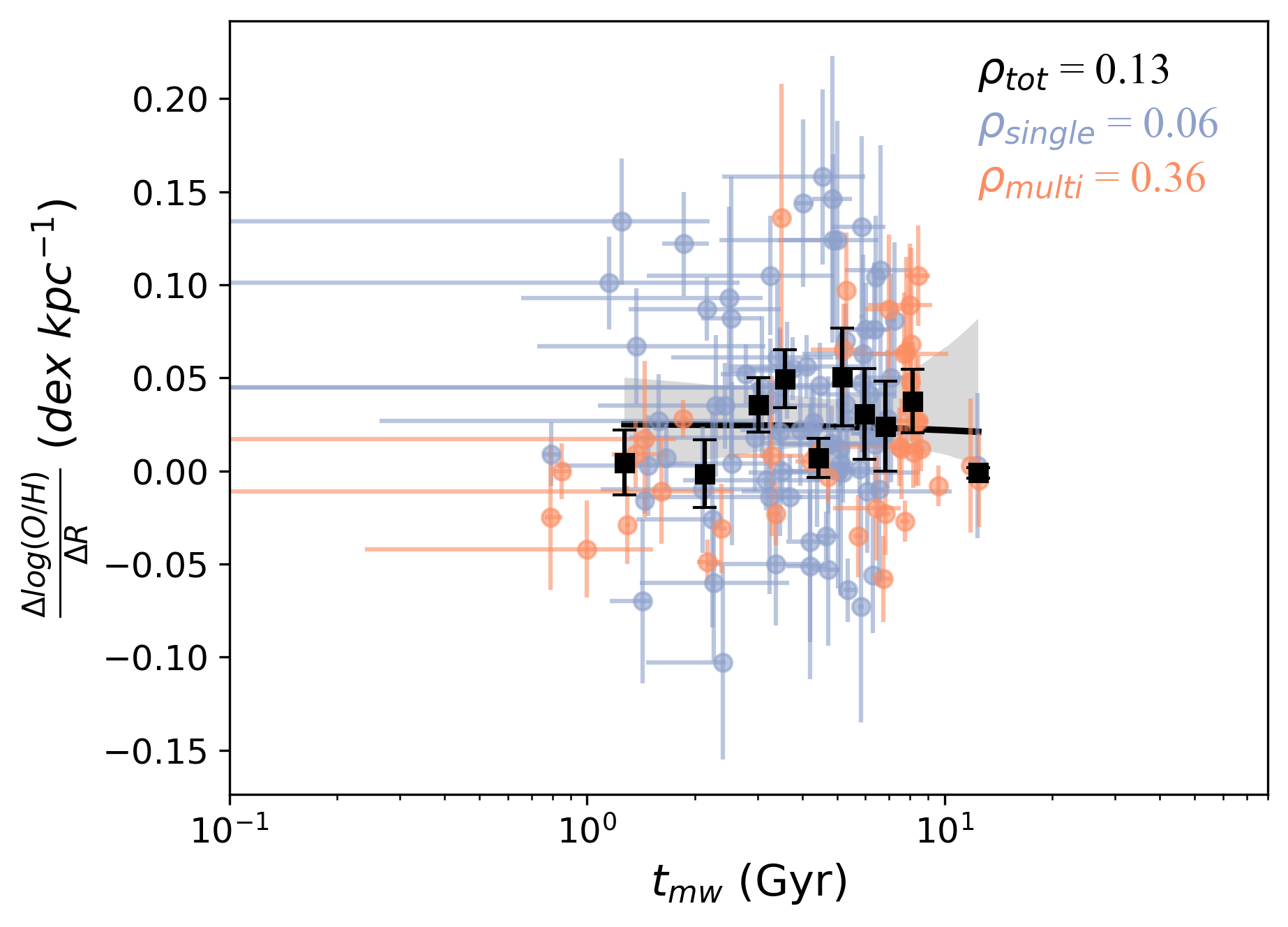}{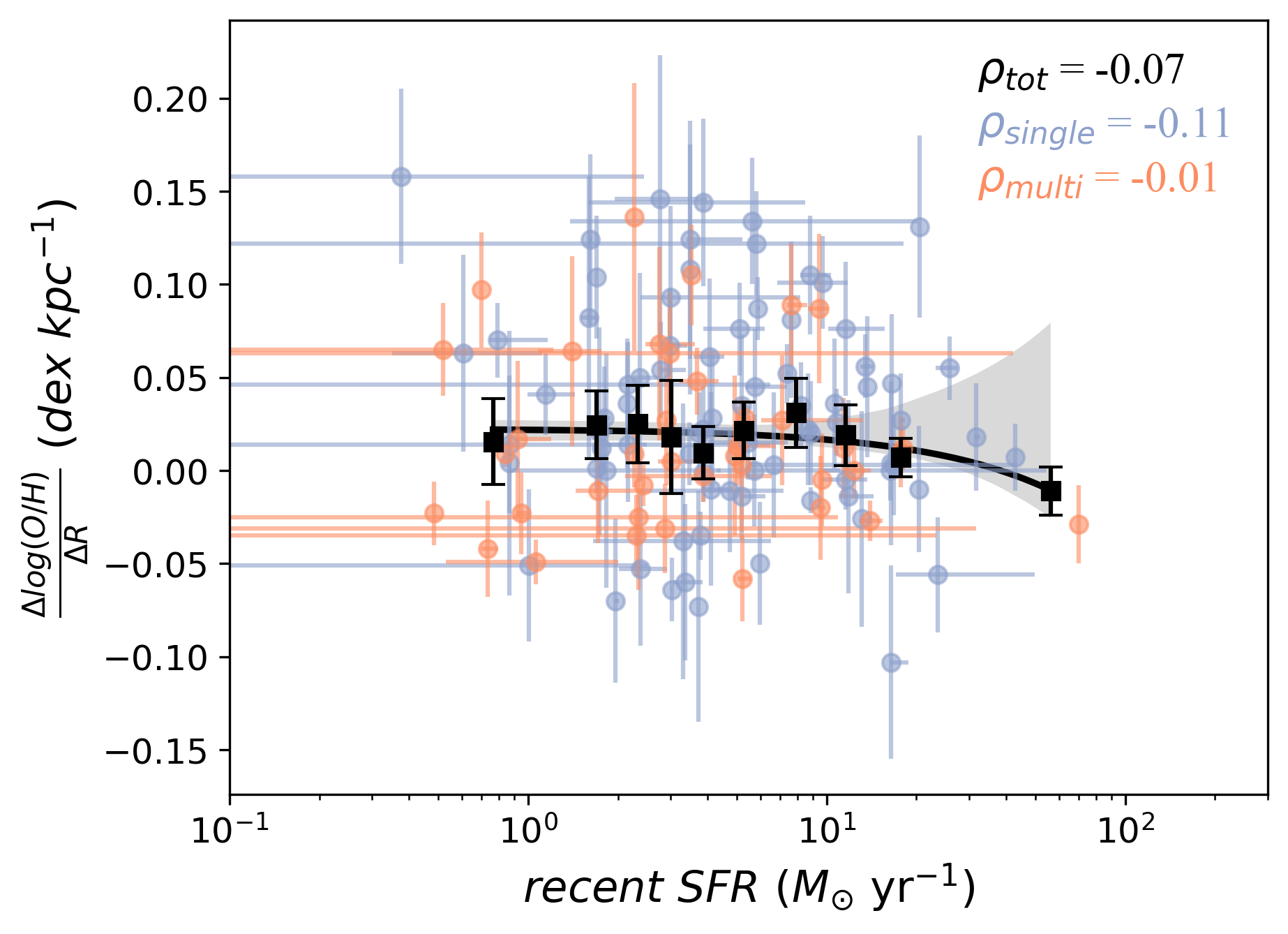}
\caption{The gas-phase metallicity gradients as a function of mass-weighted age (left) and the recent star formation rate (right). Blue represents galaxies with only one main SF episode in their SFH, orange represents galaxies with multiple SF episodes. The binned scatter plot with $1\sigma$ error-bars are plotted in black. The black solid line shows the best linear fit, and the grey shaded region shows the $1\sigma$ fitting error. The text label shows the Spearman's rank correlation coefficient for the total sample (black, $\rho_{tot}$), the single-SF sample (blue, $\rho_{single}$) and the multi-SF sample (orange, $\rho_{multi}$). No strong correlation is found in both cases.} \label{fig:mw}
\vspace{1cm}
\end{figure*}


After running $Prospector$ on this final sample, we have recorded the best-fit 9-bin SFH, the stellar mass and star formation rate in the most recent bin ($rSFR$), i.e. the mass and SFR at the time of observation, as well as the mass-weighted age ($t_{mw}$) and the dust obscuration. Figure \ref{fig:spec} shows the best-fit SED for one of the galaxies as a demonstration, while Figure \ref{fig:stack} shows the individual plots of the SFH for each of the 119 sample galaxies. While all the galaxies are actively forming stars, there is a big variety in the individual shape of the SFH. To identify galaxies with star-formation rejuvenation, we visually divide the sample into two subsets based on the number of major episodes of star formation in their SFH. Among the total selected sample, we visually identify 39 galaxies with prominent multiple star formation episodes in their SFH, i.e. an overall non-monotonic function, and 80 with only one main star formation episode, i.e. an overall monotonic function. Figure \ref{fig:ex} demonstrates the `typical' SFH curves for galaxies with single and multiple star formation episode. In the following analysis, we will test for possible correlations using both the total sample and each of these two subsets.

\section{SFH vs. metallicity gradients} \label{sec:sfh}

As shown in Figure \ref{fig:mw}, there is no obvious correlation between the gas-phase metallicity gradient and $t_{mw}$ or $rSFR$. 
To reduce high-frequency noise and identify more clearly possible trends, we have binned the points by dividing the X axis into 10 equally-sized bins and calculated the median Y values in each bin. The binned data points as well as the best linear fit are shown in black, while the grey shaded region shows the $1\sigma$ fitting error. To quantify the significance of the possible correlation, we calculate the Spearman's rank correlation coefficient for each case. Except that $t_{mw}$ shows a very weak correlation with the metallicity gradients for galaxies with multiple SF episodes (Figure \ref{fig:mw}, left panel), the sense being that for these galaxies the older means more positive gradient, no correlation is found in any of the other cases (with $|$Spearman's rank correlation coefficients$|<0.2$). In terms of $rSFR$, there is no difference between galaxies with single or multiple star formation episodes.

\subsection{Characterizing the SFH}
It is customary to use some characteristic redshift as a parameter to describe in a synthetic form the shape of a galaxy's SFH, for example the redshifts when a certain percent of the stellar mass has formed. 
Here we use an additional set of parameters, proposed by \citet{2022arXiv220402414J}, designed to provide quantitative information on the overall shape of the SFH. The four parameters used here are listed below, where $t_{obs}$ refers to the age of the Universe at the observed redshift.
$\\$
\begin{itemize}
    \item{$\tau_{tot}$}: the total time length of star-formation ($0 < t < t_{obs}$)
    \item{$\tau_{1}$}: the time length of the late period of star formation ($t_{mw} < t < t_{obs}$), referred to as $\tau_{SF}$ in \citealt{2022arXiv220402414J}
    \item{$\tau_{2}$}: the time length of the early period of star formation ($0 < t < t_{mw}$), referred to as $\tau_{Q}$ in \citealt{2022arXiv220402414J}
    \item{$Skewness$}: the third standardized moment of stellar ages ($\frac{8}{\tau_{\text {tot }}^3} \frac{\sum_{i=1}^9\left(t_i-t_{\mathrm{age}}\right)^3 \cdot \mathrm{SFR}_{\mathrm{i}} \delta \mathrm{t}_{\mathrm{i}}}{M_*}$)
\end{itemize}

We refer the readers to Section 2.4 of \citet{2022arXiv220402414J} for detailed descriptions of these parameters. The ratio between $\tau_{1}$ and $\tau_{2}$, which strongly correlated with $Skewness$, serves as a good metric for quantifying the asymmetry of SFHs. 
We note that galaxies that fall at the extreme ends of the $Skewness$ range usually have multiple SF episodes, i.e. a more complex SFHs than one characterized by an increasing phase, a peak more or less extended, and a decrease. All these parameters enable us to better identify the overall shape of SFHs as well as capture some feature in the star formation history.

Finally, following Giavalisco et al. (2023, in prep.), in an effort to further characterize the SFH in physical terms, we also use a new time parameter, defined to describe the evolutionary state of galaxies at the time of observation, named the evolutionary time, $t_E$: 

\begin{equation}
    t_E=\int_{t_i}^{t_{obs}} \frac{1}{\left(t-t_i\right)} \times \frac{1}{sSFR\left(t-t_i\right)} \times dt,
\end{equation}
$$\\$$

where $sSFR$ is the specific star formation rate, $t_{obs}$ the time at observation, and $t_i$ is the time when star formation started in the galaxy, which in practice can be assumed coincident with the Big Bang in this work. The quantity inside the integral is, essentially, the mass doubling time of the galaxy at the time $t$ in unit of the Hubble Time (e.g. see \citealt{2019MNRAS.487.5416T}). Thus, $t_E$, the integral of this quantity over the life time of the galaxy at the time of observation, is the average amount of time the galaxy spent doubling its stellar mass. Approximately, galaxies with the same $t_E$ are in the same evolutionary phase. Therefore, the relation between $t_E$ and the average of galaxy properties at $t_{obs}$ provides a description on the evolution of such properties with time. 

\subsection{Relationship between metallicity gradients and the SFH}

We now investigate the relationship between the gas-phase metallicity gradients and the five SFH parameters defined above.

\begin{figure*}
\plottwo{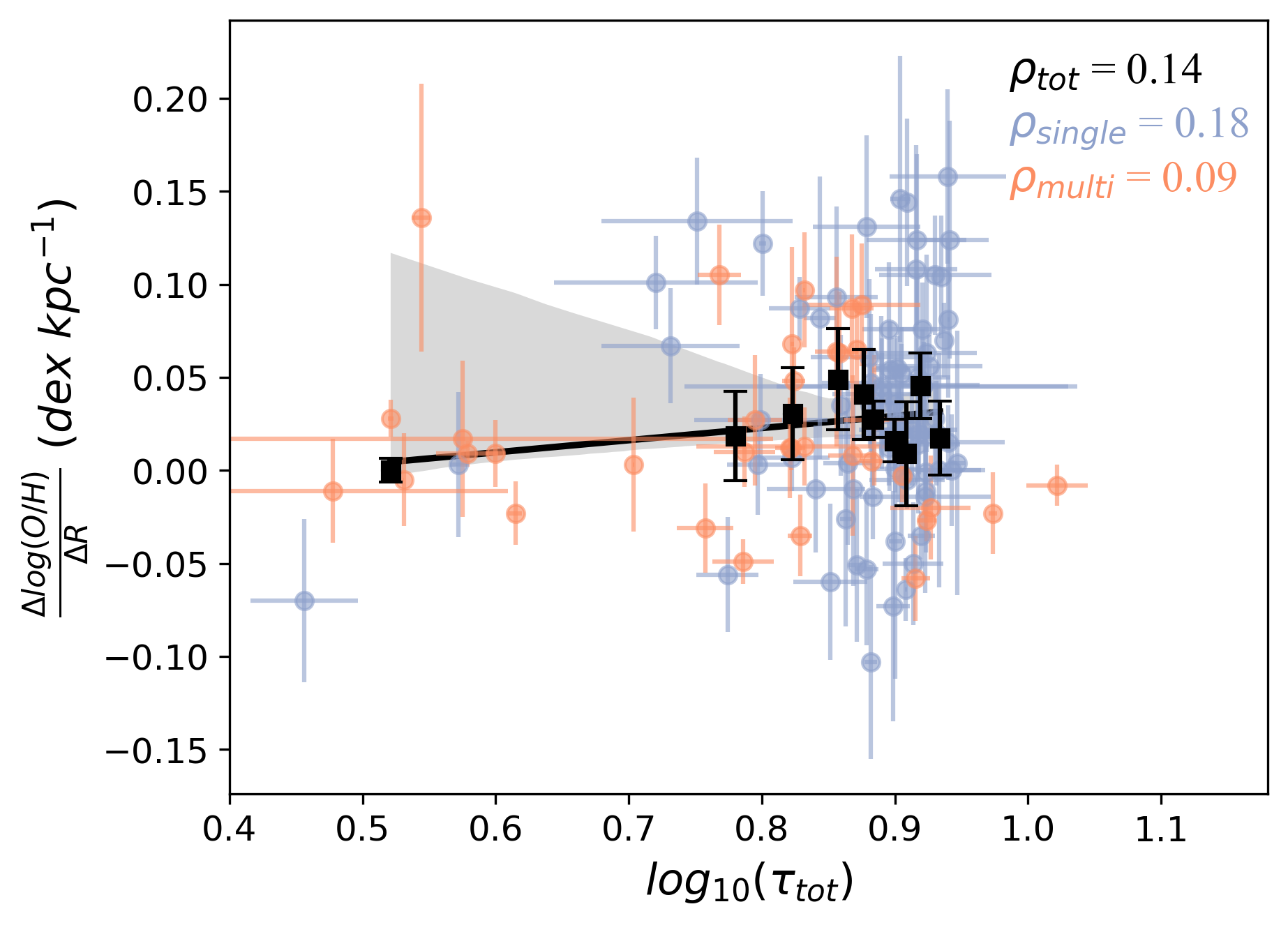}{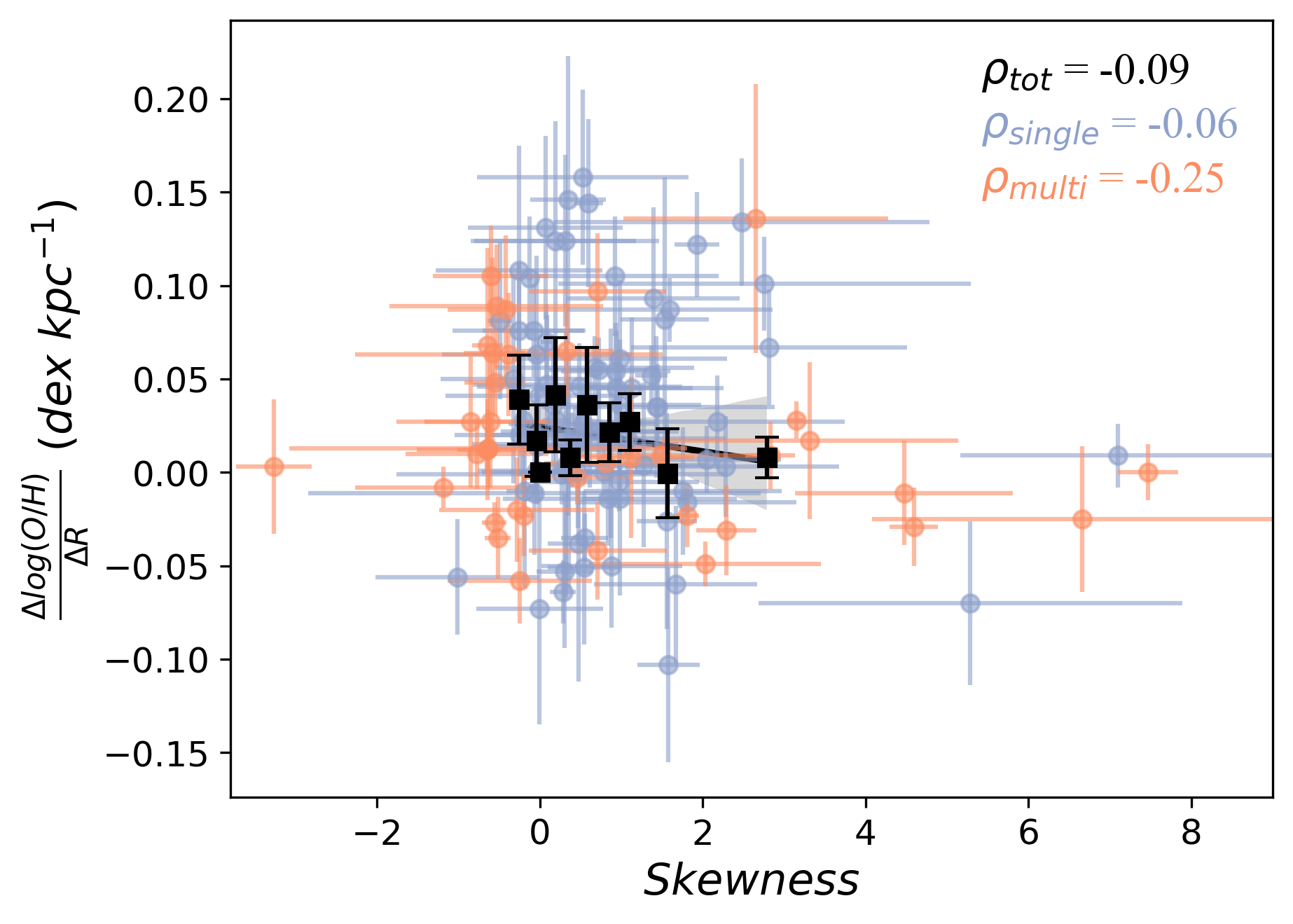}
\plottwo{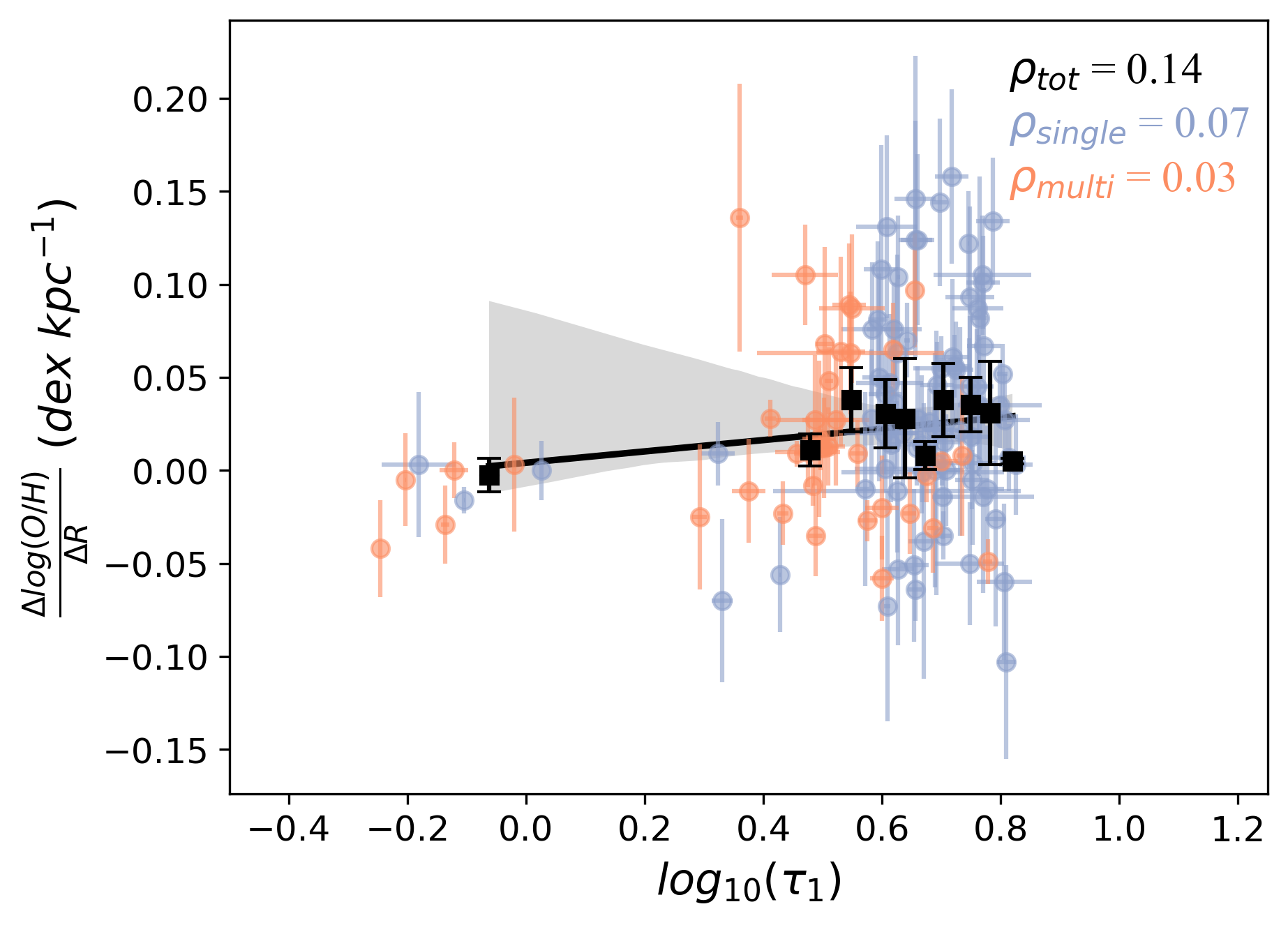}{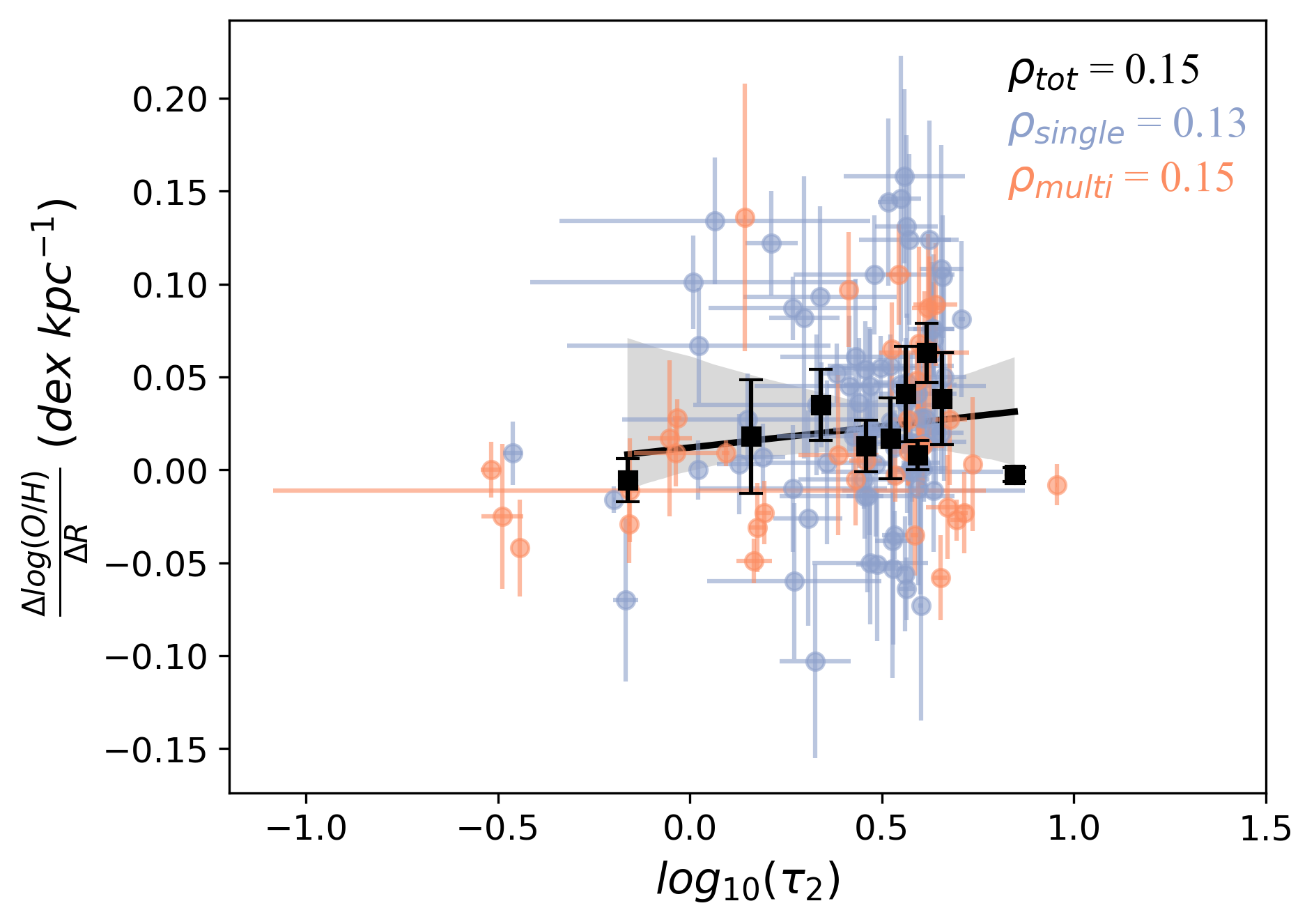}
\caption{The relationship between the gas-phase metallicity gradients and four SFH characteristics. Marks and labels are the same as Figure \ref{fig:mw}. According to the low Spearman's coefficients, no strong correlations are identified here. \label{fig:tau}}
\vspace{1cm}
\end{figure*}

As shown in Figure \ref{fig:tau}, with all the Spearman's rank correlation coefficients less than 0.2, we find no clear correlation between any of the four parameters that describe the overall shape of the SFH and the metallicity gradients. There is a hint that galaxies with single or multiple star formation episodes might occupy different regions in the diagrams of $\tau_1$ and $\tau_2$ versus the gradients. However, when we look at the $Skewness$ (analogous to the ratio $\tau_1$/$\tau_2$), the distribution of points shows no dependence on the number of SF episodes. Similarly, the relationship between $\tau_{tot}$ and the metallicity gradients is also insensitive to the number of SF episodes. The results are summarized in Table \ref{tab:data}, the machine-readable version will be available online.

\startlongtable
\begin{deluxetable*}{lcccccccccc}
\tabletypesize{\scriptsize}
\tablewidth{0pt}
\tablecaption{The galaxy properties and SFH characteristics derived from SED fittings \label{tab:data}}
\tablehead{
\colhead{Field} & \colhead{ID}& \colhead{Redshift} & \colhead{lg($M/M_{\odot}$)} & \colhead{Z slope(dex/kpc)} & \colhead{rSFR($M_{\odot}$/yr)} & \colhead{$t_{mw}$(Gyr)} & \colhead{$\tau_{tot}$(Gyr)} & \colhead{$\tau_1$(Gyr)} & \colhead{$\tau_2$(Gyr)} & \colhead{Skewness} \\
}
\colnumbers
\startdata 
GOODSN  &7717  &1.51  &9.102  &0.124  &1.613  &4.869  &8.247  &4.567  &3.724  &0.311 \\
GOODSN  &7904  &1.02  &8.723  &0.063  &0.608  &5.92  &8.387  &4.219  &4.306  &-0.035 \\
GOODSN  &8344  &1.24  &9.257  &0.035  &8.221  &2.435  &7.612  &6.287  &2.198  &1.453 \\
GOODSN  &8797  &1.98  &9.929  &0.013  &5.043  &7.457  &6.793  &3.228  &4.032  &-0.635 \\
GOODSN  &9041  &0.88  &8.349  &0.158  &0.376  &4.563  &8.705  &5.221  &3.617  &0.527 \\
GOODSN  &9558  &1.43  &9.111  &0.046  &2.161  &4.47  &8.346  &4.927  &3.531  &0.488 \\
\enddata
\tablecomments{The IDs are matched to the CANDLES photometric catalog. The metallicity gradient measurements are taken from \citet{2021ApJ...923..203S}. Only part of the table is shown here, the full data will be available online alongside publication.}
\end{deluxetable*}

\begin{figure}
\plotone{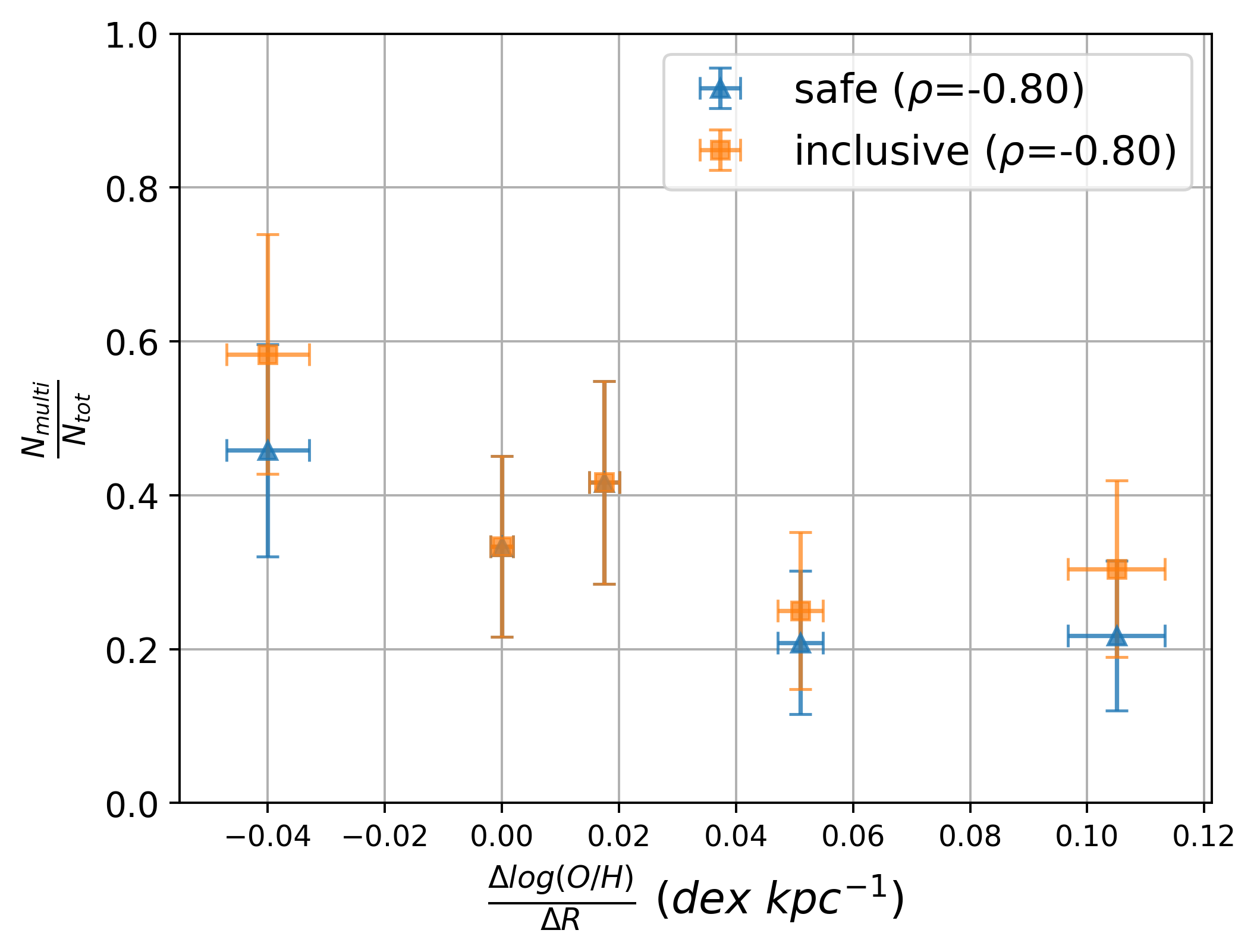}
\caption{The fraction of galaxies showing multiple star-forming episodes in their reconstructed SFH at different gas-phase metallicity gradients. The blue points show the safe selection of multi-SF galaxies, while the orange points show the more inclusive selection. An obvious anti-correlation is seen in each of the case. }\label{fig:mul}
\end{figure}

To have additional insight into the possible effects of multiple episodes of star formation during the SFH of galaxies on the gas-phase metallicity gradients, in Figure \ref{fig:mul} we plot the fraction of galaxies whose SFH shows evidence of distinct and repeated episodes of star formation versus the gradients. Such a fraction is defined as the ratio of the number of galaxies with multiple SF episodes to the total number of galaxies, and to do so we adopt the same selection criteria of `multiple SF episodes' as described in \citet{2022arXiv220804325J}. To test for the effect of uncertainties in such a selection, we also develop a more inclusive sample of multi-SF galaxies by adding in all `possible' cases, whenever the error bars in the reconstructed SFH make it difficult to count the number of star-formation episodes (see Fig. 11 of \citealt{2022arXiv220804325J}). As shown in the figure, for both the original `safe' selection and the `more inclusive' one a highly-significant correlation of the multi-SF fraction as the metallicity gradient increases is observed, with a Spearman's coefficient of -0.80 and a p-value of 0.10. Therefore, galaxies with more negative gas-phase metallicity gradients are more likely to have experienced multiple SF episodes in their SFH, and this effect seems robust with the details of the selection of SFH with multiple SF episodes.

\begin{figure*}
\plottwo{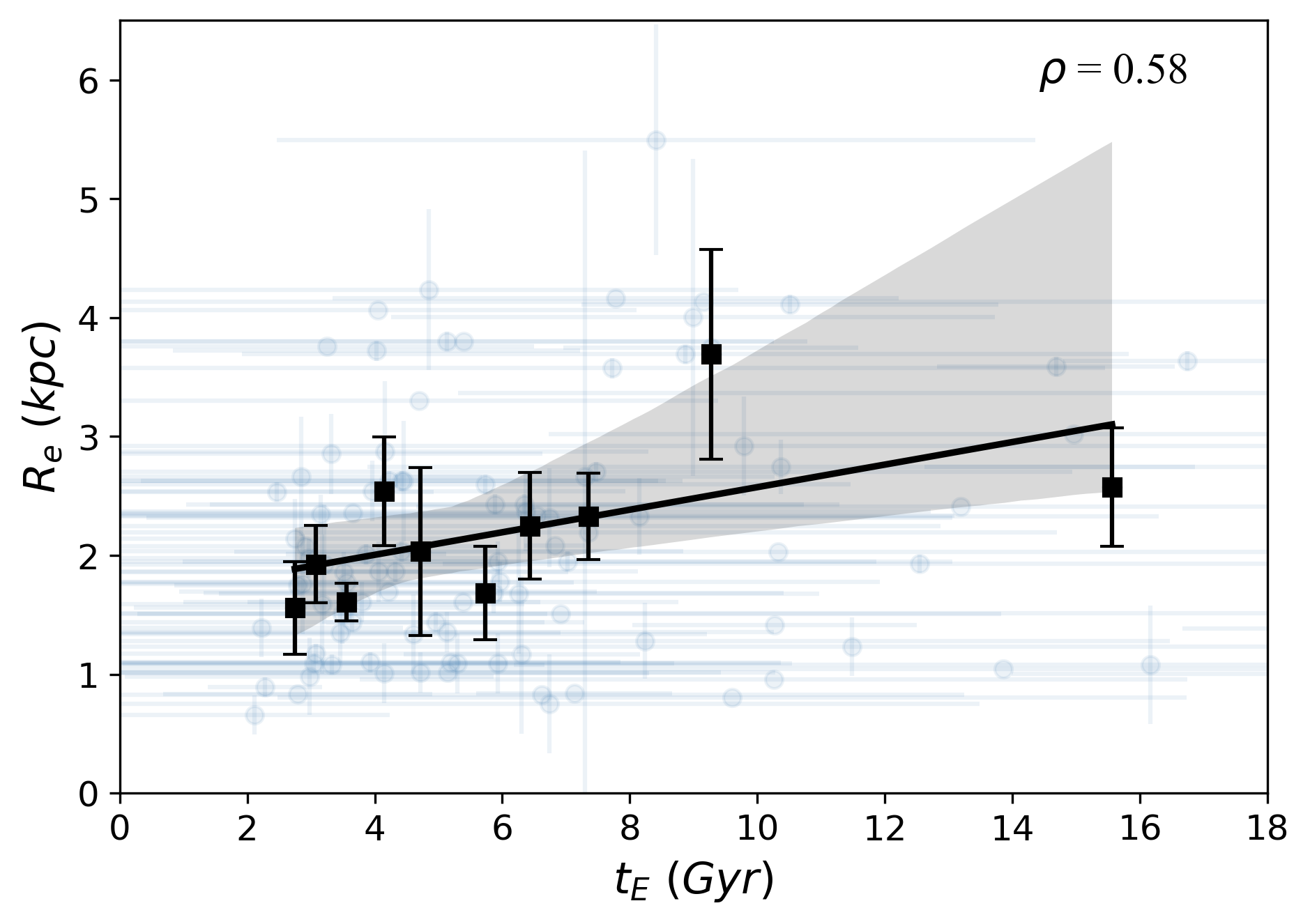}{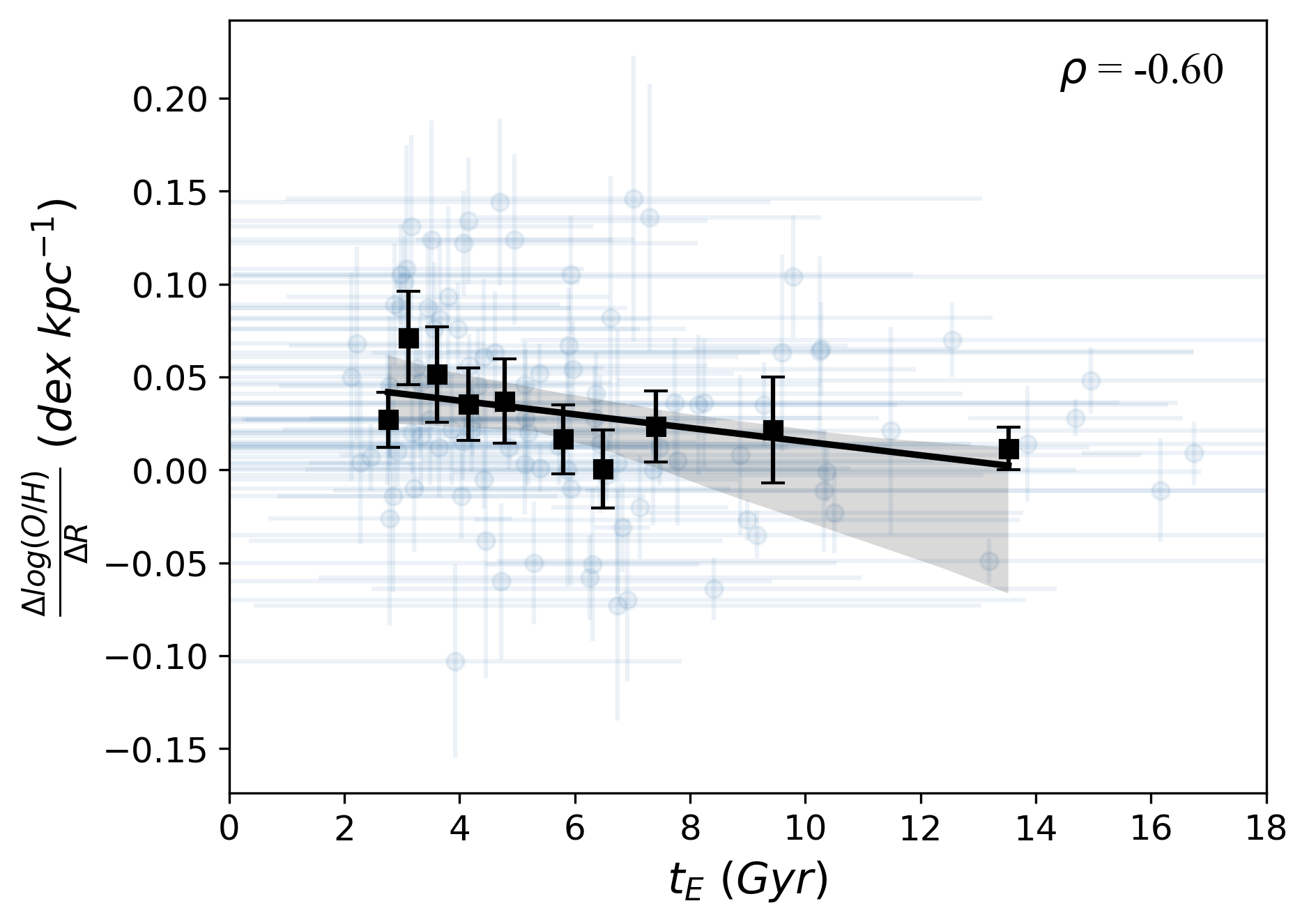}
\caption{The effective radius $R_e$ (left) and the gas-phase metallicity gradients (right) as a function of the evolution time $t_E$. The binned scatter points are shown in black squares with $1\sigma$ error-bars. The black solid line shows the best linear fit, and the grey shaded region shows the $1\sigma$ fitting error. The unbinned data points are plotted in light blue. \label{fig:te}}
\end{figure*}

We then investigate the relationship between the gas-phase metallicity gradients and the evolutionary time $t_E$, which is directly calculated from the SFH by replacing the integral with the sum over the 9 bins of look-back time. Figure \ref{fig:te} shows the effective radius $R_e$ and the metallicity gradients as a function of $t_E$ in Gyr. The size of these galaxies increases with the evolutionary time with a Spearman's coefficient of 0.58 and a p-value of 0.03, inferring that star-forming galaxies grow larger along their evolution process. This finding is consistent with previous work about the size evolution of star-forming galaxies (e.g. \citealt{2017ApJ...834L..11A, 2019A&A...625A.114J}). The metallicity gradient is also significantly correlated with $t_E$, with a Spearman's coefficient of -0.60 and a p-value of 0.01. The gradient goes from positive values at small $t_E$ to $\approx 0$ at large $t_E$. The meaning of this correlation is that less evolved galaxies tend to have more positive metallicity gradient. In time, these gradients become progressively smaller and smaller, and possibly negative, as the galaxies evolve. 

\begin{figure*}
\plottwo{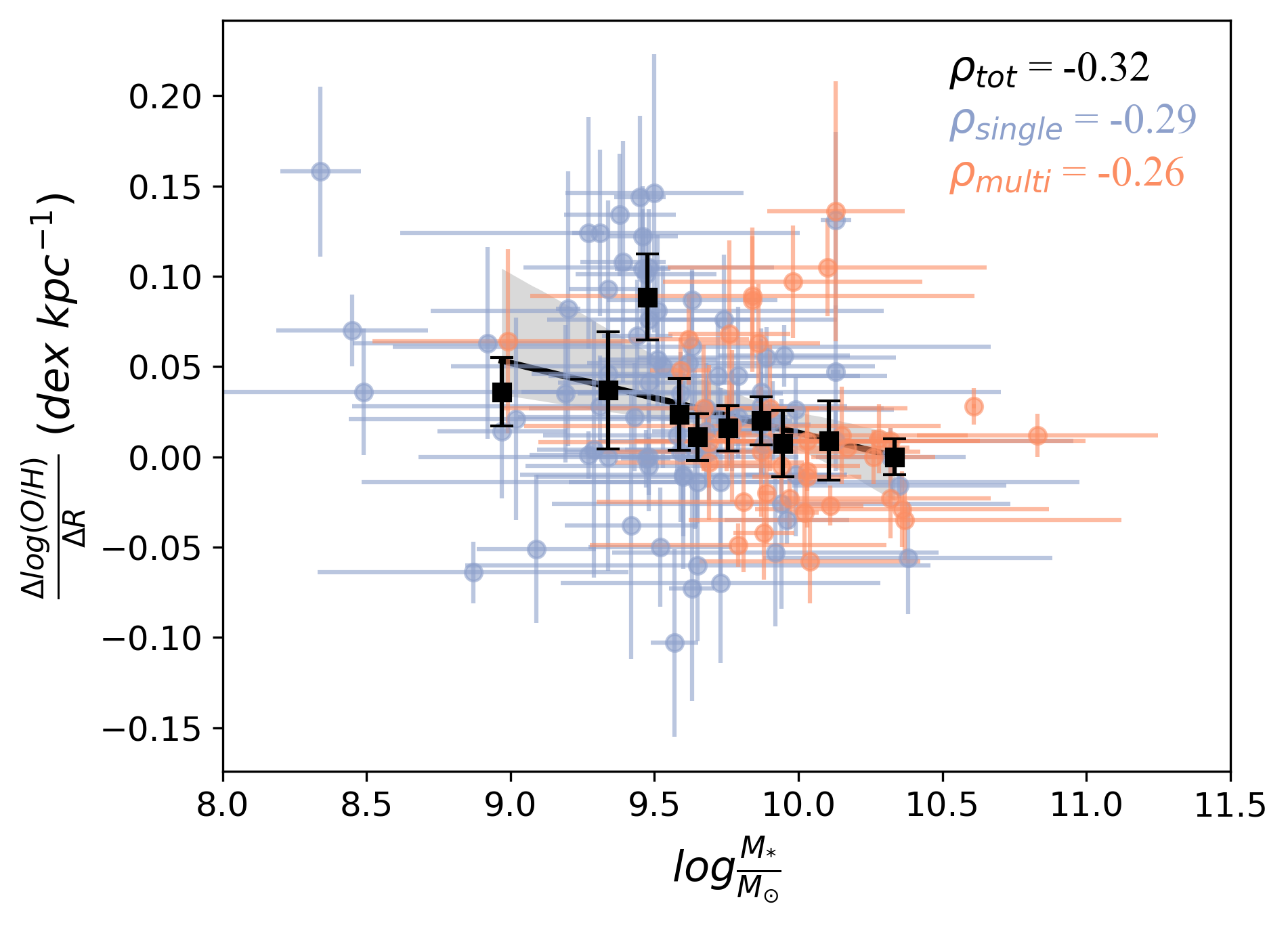}{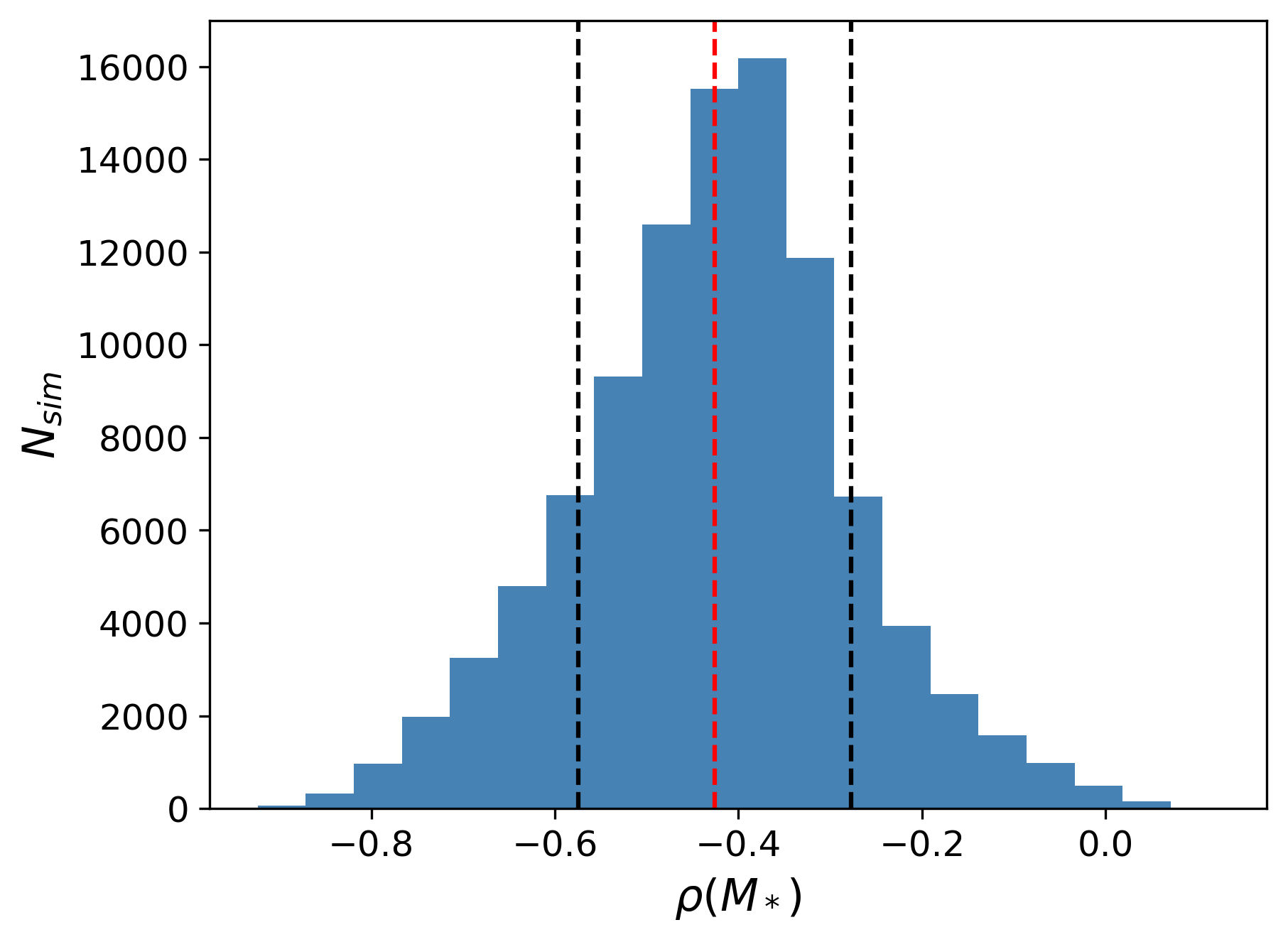}
\caption{The gas-phase metallicity gradients as a function of the stellar mass. The best linear fit as well as the 1$\sigma$ error range of the binned scatter plot are shown in black. The number distribution of the correlation coefficient after bootstrapping is shown in the right panel. The red and black dashed lines show the center and the 68\% confidence interval of the distribution. \label{fig:mass_mul}}
\end{figure*}

To gain further insight into the physical meaning of these correlations, in the left panel of Figure \ref{fig:mass_mul}, we plot the gas-phase metallicity gradient versus the stellar mass and color-code the points for having one major SF episode (blue) or multiple episodes (orange) in their SFH. Although with large scatter, there is a significant correlation ($\rho=-0.32$ with a p-value $<0.01$) such that smaller galaxies have positive gradients and more massive galaxies mildly negative ones, with galaxies with $log~M_*/M_{\odot}\approx 10$ having flat gradients. Given the big uncertainties in the measurement of $\rho$, we calculate the bootstrap distribution of $\rho$ by re-sampling the data with corresponding error bars 100,000 times. As shown in the right panel of Figure \ref{fig:mass_mul}, the $1-\sigma$ bootstrap confidence interval is derived as [-0.57, -0.28], which confirms that the negative correlation between metallicity gradient and stellar mass is robust. Galaxies with multiple SF episodes are mostly concentrated toward the high-mass end of the distribution, while single-peak SFHs populate the low-mass side. The fact the galaxies with multiple SF peaks are on average more massive than those with one peak only has been reported by \citet{2022arXiv220804325J}. The novelty here is that the former on average shows positive gas-phase metallicity gradients, while the latter shows flat or mildly negative gradients. Thus, the picture that seems to be emerging from these analyses is that, on average, galaxies develop positive gradients early on in their evolution. As evolution proceeds, these positive gradients are progressively reduced, becoming flat or mildly negative for highly evolved galaxies. This process seems to be more effective in galaxies with multiple bursts in their star formation history. Taking all together these elements suggests that gas accretion and radial transport of gas reduce the gradients by progressively enriching the inner regions and diluting the chemical content of the outer ones with fresh gas. 

\section{Measurements of colors and attenuation} \label{sec:color}

\begin{figure}
\plotone{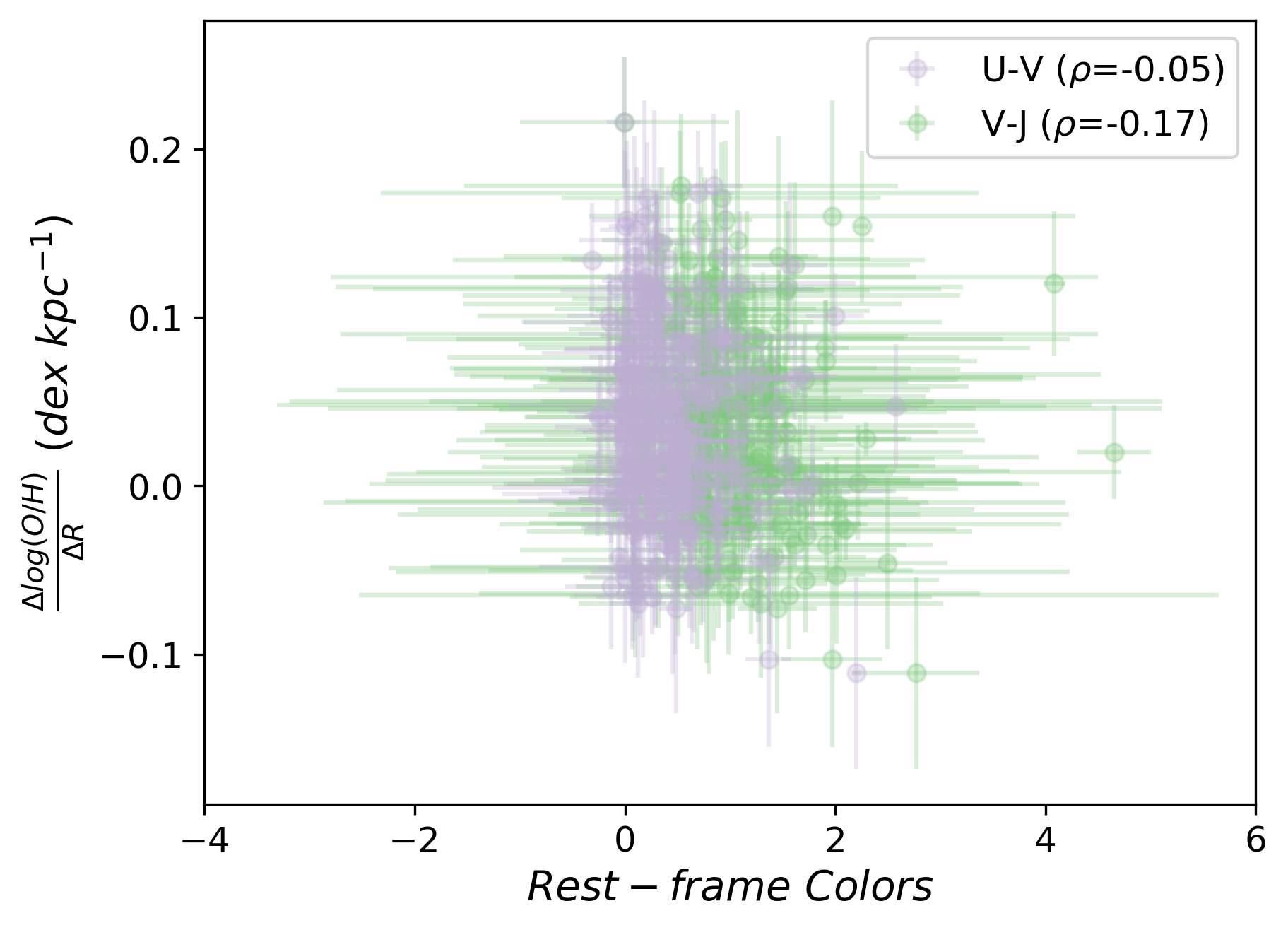}
\caption{The relation between the gas-phase metallicity gradients and the rest-frame $U-V$ and $V-J$ colors. \label{fig:uvj}}
\end{figure}

The rest-frame colors as well as the radial color gradients are closely related to the metallicity of galaxies. Recent studies at moderate and high redshifts ($z\approx2$) have shown that stellar age, dust attenuation and metallicity can all affect the color gradients of star-forming galaxies \citep[e.g,][]{2017ApJ...844L...2L, 2022ApJ...941L..37M}. In an effort to disentangle the degeneracy among these effects, we first test the relationship between the metallicity gradients and the rest-frame UV and optical colors. 
The best-fit SEDs are convolved with the Johnson/Bessel transmission filters to get the flux densities of the rest-frame U, V, J bands. As shown in Figure \ref{fig:uvj}, both the $U-V$ and $V-J$ colors show no correlation with the metallicity gradients.

\begin{figure}
\plotone{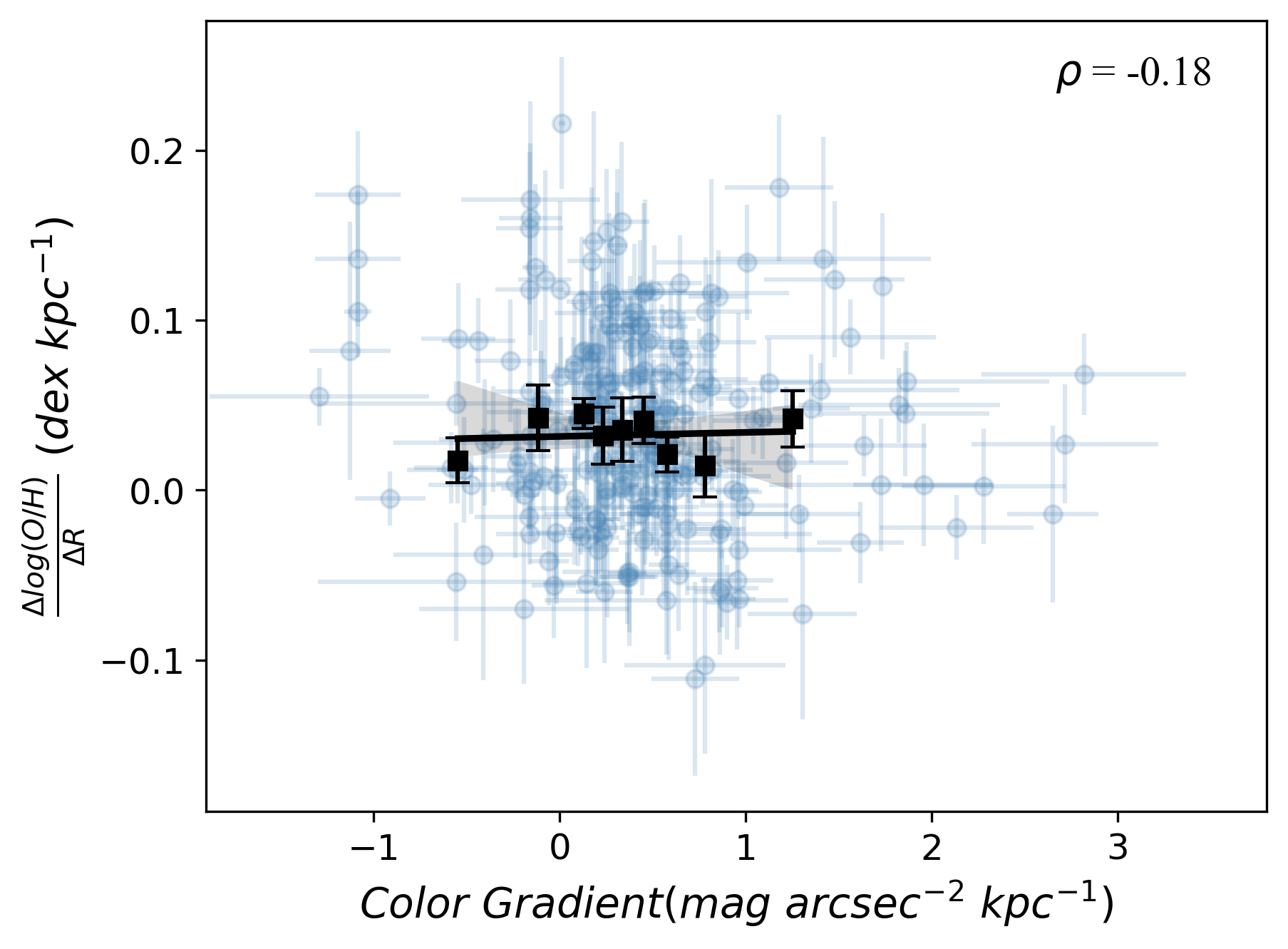}
\caption{The relation between the gas-phase metallicity gradients and the radial color gradients. Marks and labels are the same as Figure \ref{fig:te}. \label{fig:color}}
\end{figure}

We continue to derive the color gradients for galaxies in our sample. We measured the azimuth-averaged radial surface brightness profiles in the PSF-matched {\it HST}/ACS V band (F606W) and z band (F850LP) images for each galaxy using the \textit{photutils} package \citep{2020zndo...4044744B}, corresponding approximately to the $UV-U$ color in the rest frame. We then performed a linear fit to characterize the color profiles between these two bands in semi-log scale (e.g.,\citealt{2005ApJ...622..244W, 2016ApJ...822L..25L}). We adopted an inner radius limit of 0.3 kpc to reduce the PSF effect, and an outer radius cut at 2$R_e$. Then we expressed the color gradient as $d(F(V)-F(z))/dlog(R)$ (0.3 kpc $<$ R $<$ 2$R_{e}$).

The derived color gradients of our galaxies are generally flat, with some scattering on both positive and negative sides (Figure \ref{fig:color}). We calculated the Spearman's rank coefficient of the correlation between the color gradients and metallicity gradients, finding no significant correlation between these two quantities. This suggests that the gas-phase metallicity gradient does not drive color gradients in these galaxies. 

We have also considered the possibility that  differential dust attenuation across the profile of galaxies could also give rise to color gradients and in turn, provide systematic bias in the measures of  metallicity gradients. As discussed by \citet{2021ApJ...923..203S}, the measures are obtained from spectral indexes based on the ratio of multiple emission lines located inside a well defined range of wavelength. Using Monte Carlo simulations applied on the line maps, they showed that a spatially varying extinction, if any, could only change the metallicity gradient measurements by 5-30\%, depending on the diagnostic and the intrinsic metallicity gradient. Furthermore, according to the 3D-HST Balmer-line survey, the $A_V$ profiles in the redshift and mass range covered by our sample ($\log~M_*/M_{\odot}<10.5$) are flat outside the inner $\sim$1 kpc \citep{2016ApJ...817L...9N}, implying that the dust extinction gradients of the galaxies in our sample should also be, on average, flat.

\begin{figure*}
\plottwo{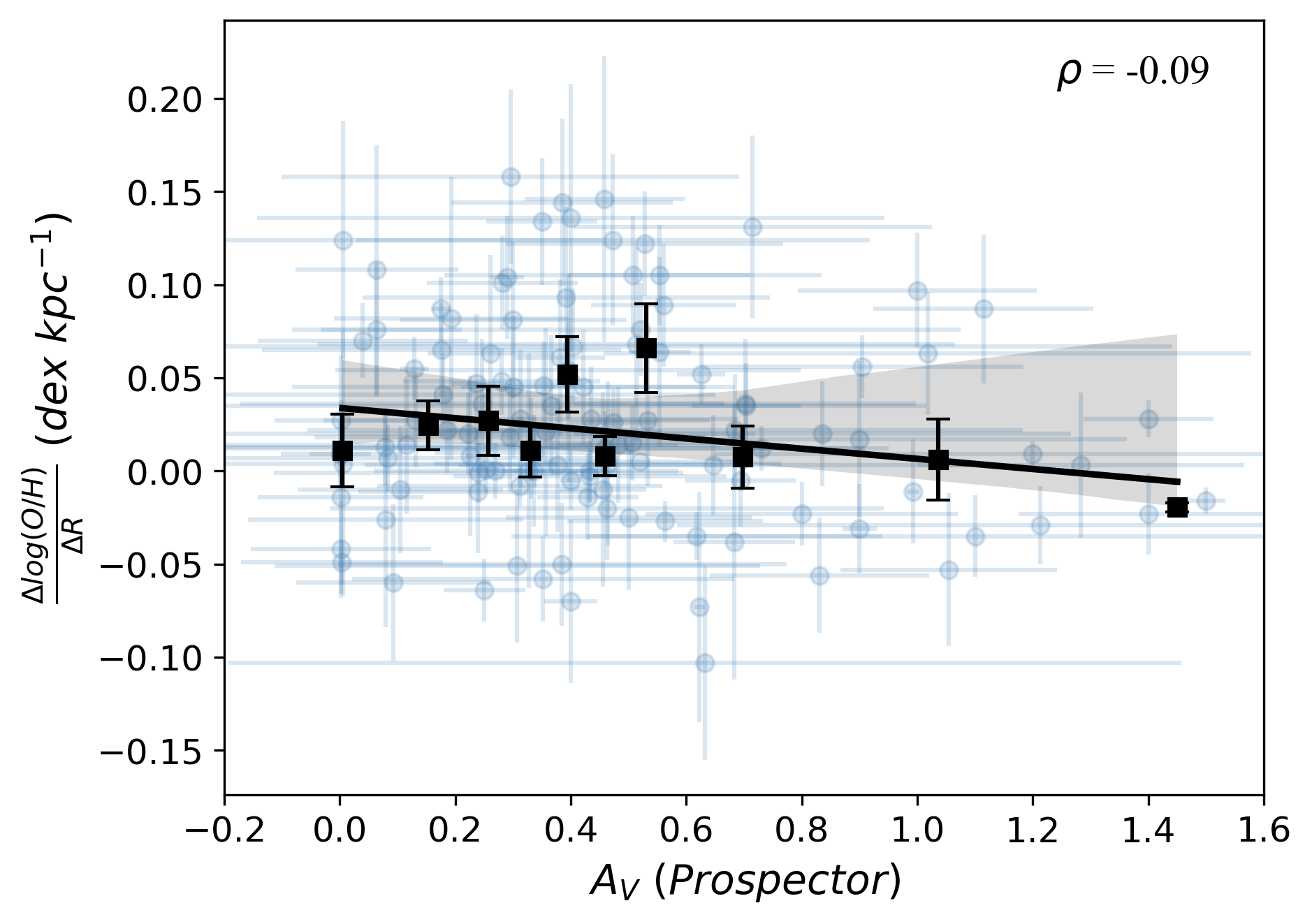}{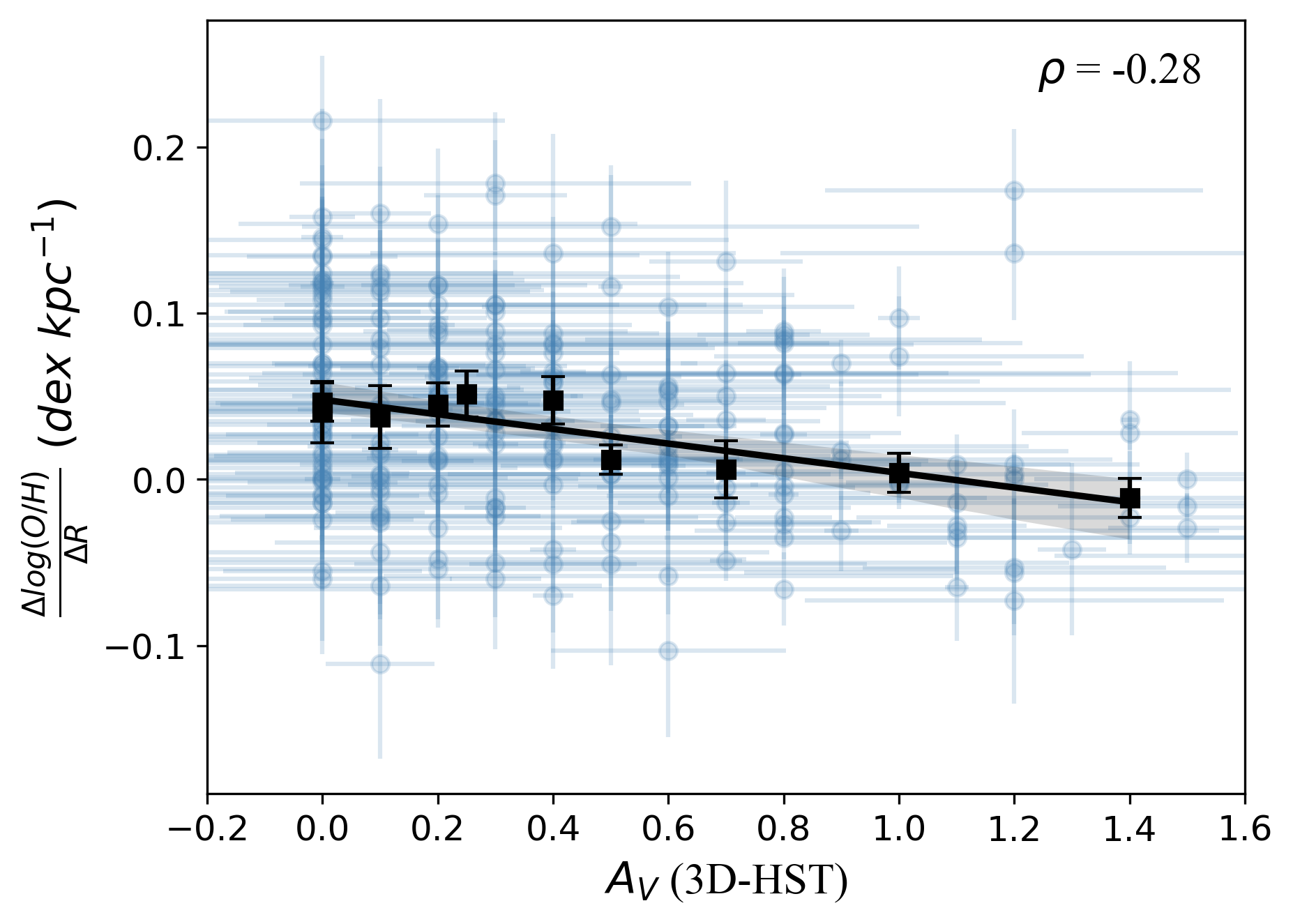}
\caption{The relation between the gas-phase metallicity gradients and the V band attenuation $A_V$. Marks and labels are the same as Figure \ref{fig:te}. In the left panel, $A_V$ is derived from SED fittings using $Prospector$; in the right panel, $A_V$ is directly taken from the 3D-HST data release (only reported to one significant digit). \label{fig:av}}
\end{figure*}

To test the robustness of the measures of $A_V$, we have used both the outputs from our $Prospector$ SED fittings and the measures from the 3D-HST catalogs  \citep{2014ApJS..214...24S}, which have been obtained with a different procedure that uses the FAST package\citep{2009ApJ...700..221K}, assuming discrete grid values of $A_V$. The results are compared in Figure \ref{fig:av}. As seen in the figure, the two results are qualitatively consistent with each other, possibly showing a tentative, very weak anti-correlation. The Spearman's coefficient for the Prospector measure is $\rho\sim-0.1$ while that for the 3D-HST measures is $\rho\sim-0.3$ with a p-value $<0.01$, suggesting a more significant, although very week anti-correlation. In any case, both plots suggest that more dusty galaxies are more likely to develop steeper (more negative) metallicity gradients, in qualitative agreement with the general idea that more evolved galaxies develop more negative gradients. 

Finally, we conducted two additional tests of the robustness of the results above. In the first test, with the results from the SED modeling in hand, we looked for any possible dependence of the gradients on the dust attenuation ($A_V$) on the distribution of metallicity gradients, which could be evidence that the gradients themselves are the results of the dust-metallicity degeneracy in the SED fitting procedure. We split the sample into two equal-size bins in $A_V$ measurements (low attenuation bin vs. high attenuation bin) and studied the distribution of the gradients in each bin. A K-S test shows that the distributions of metallicity gradients in the two bins are statistically equivalent. This is not surprising, given the overall rather low values of $A_V$ in our sample. In the second test, we studied the dependence of the gradients with the size of the galaxies, looking for the possibility that stronger gradients could be found in larger galaxies simply because of the larger dynamic range of radii. We similarly split the sample into two equal-size bins of $R_e$ and studied the distributions of gradients in both bins, finding no significant difference between them. In conclusion, the metallicity gradient measurements discussed here are not significantly biased by the gradients in the dust extinction or size measurements.

\section{Galaxy morphology vs. metallicity gradients} \label{sec:morph}
\begin{figure}
\plotone{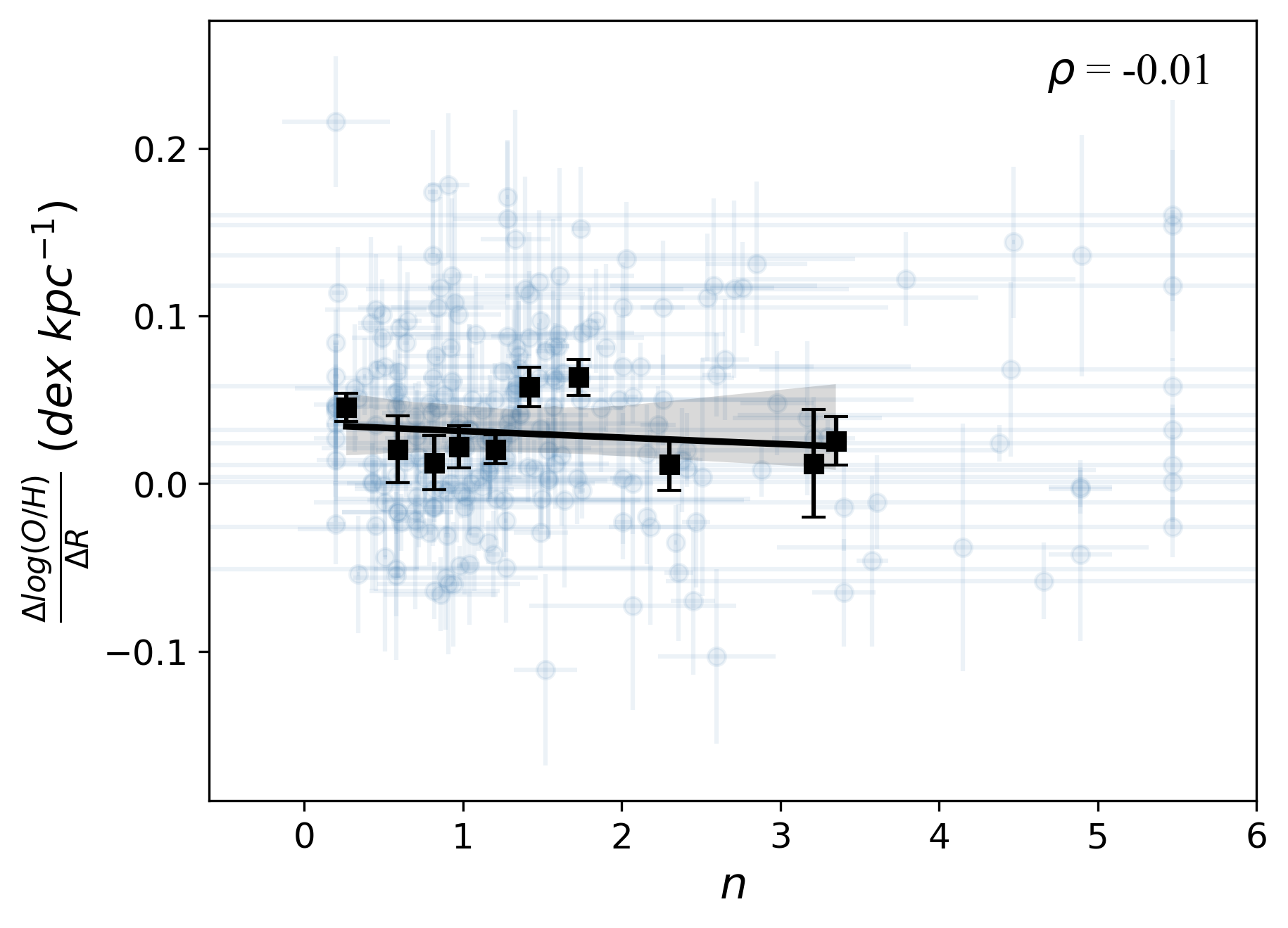}
\caption{The binned scatter plot between the gas-phase metallicity gradients and the S\'ersic indexes. Marks and labels are the same as Figure \ref{fig:te}.\label{fig:n}}
\end{figure}

Another galaxy property that may be related to metallicity gradients is the morphology. We characterize the morphology of our sample of 238 galaxies using a number of diagnostics. We adopt the S\'ersic parameters $n$ and $R_e$ provided by \citet{2012ApJS..203...24V}. As shown in Figure \ref{fig:n}, the relationship between gas-phase metallicity gradient and the S\'ersic index $n$ is flat.

\begin{figure*}
\plottwo{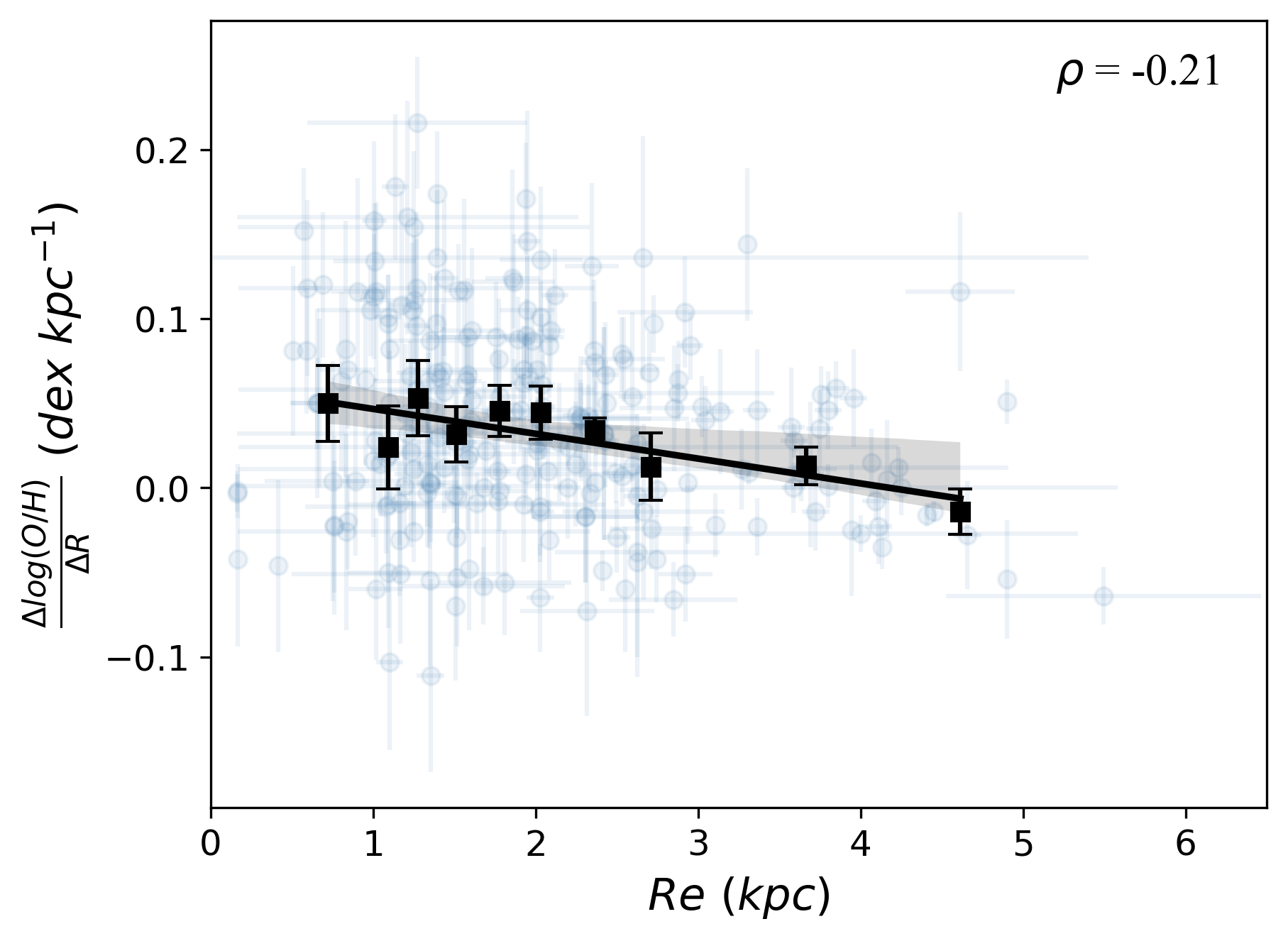}{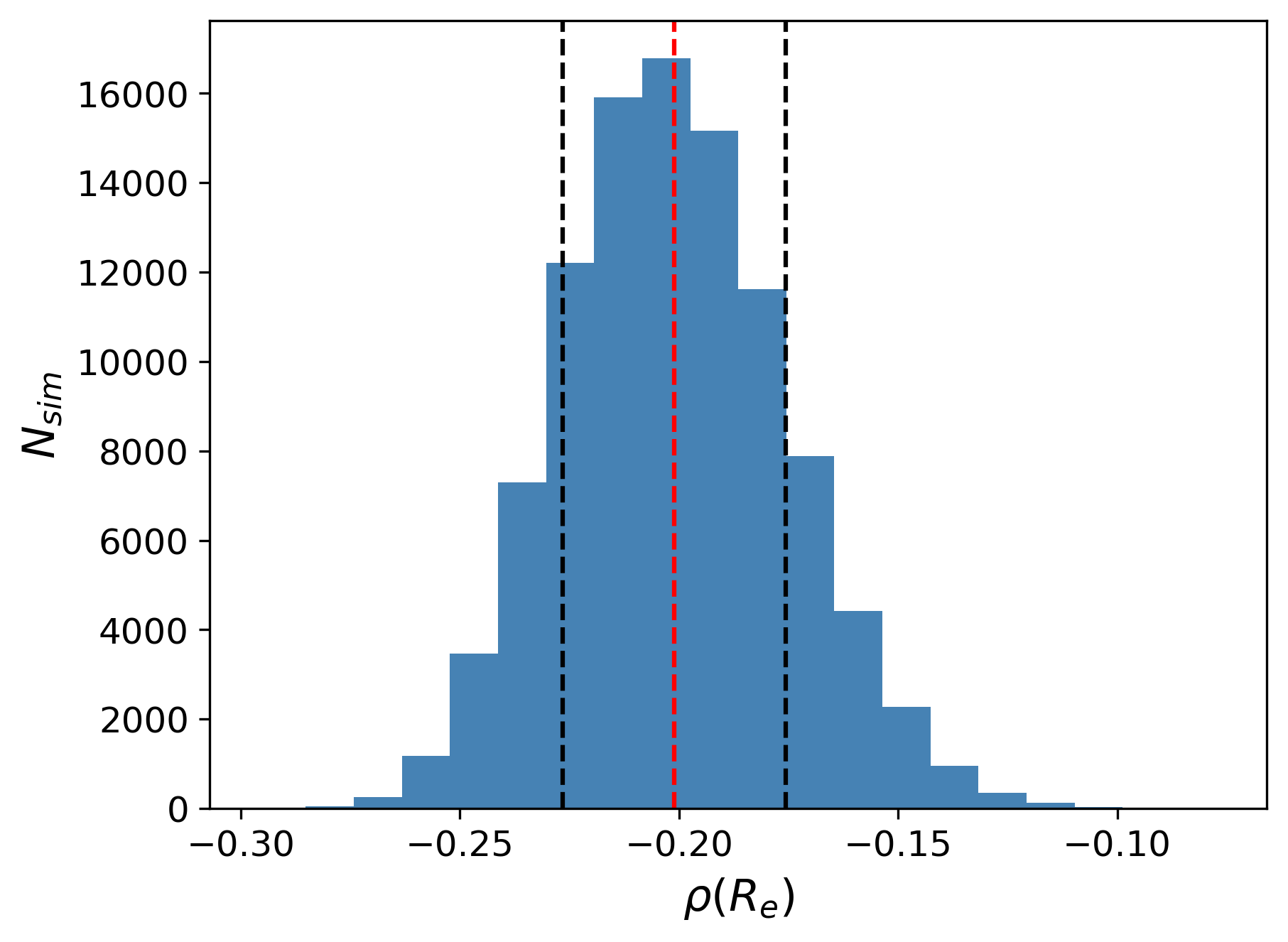}
\caption{The binned scatter plot between the gas-phase metallicity gradients and the effective radius $R_e$. The number distribution of the correlation coefficient after bootstrapping is shown in the right panel. The red and black dashed lines show the center and the 68\% confidence interval of the distribution.  \label{fig:re}}
\end{figure*}

In the top right panel of Figure 8 in \citet{2021ApJ...923..203S}, they show the correlation between gas-phase metallicity gradient and the circularized effective radius normalized by the population average at the mass and redshift of the galaxy, reporting null correlations between these two measurements. As shown in Figure \ref{fig:re}, we revisit this test and find consistent results. The 68\% confidence interval of the bootstrap distribution gives $\rho\in[-0.23, -0.18]$. Though very extended galaxies usually show more negative metallicity gradients, there is no correlation within $R_e\sim3~kpc$. 

\begin{figure*}
\plottwo{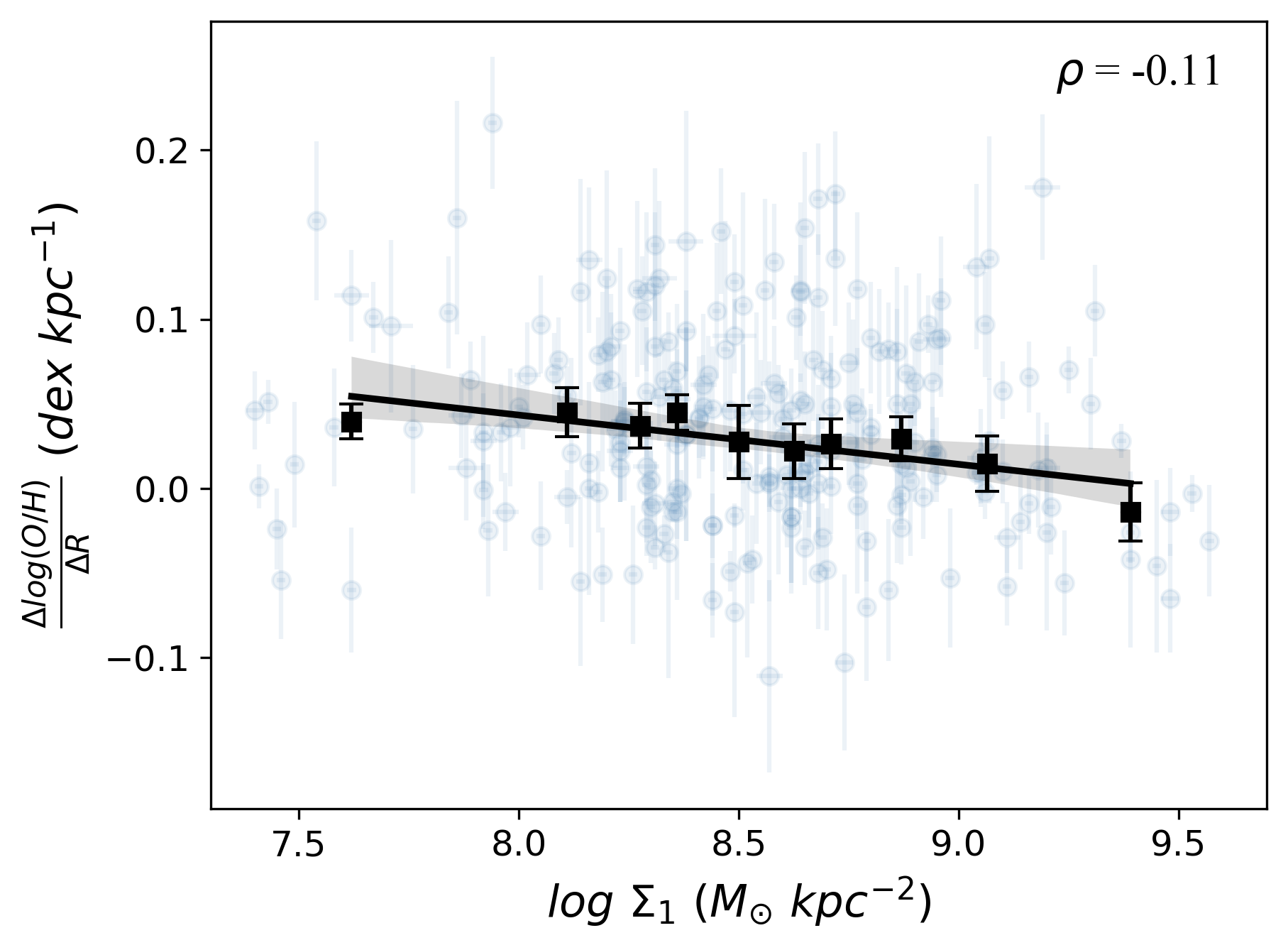}{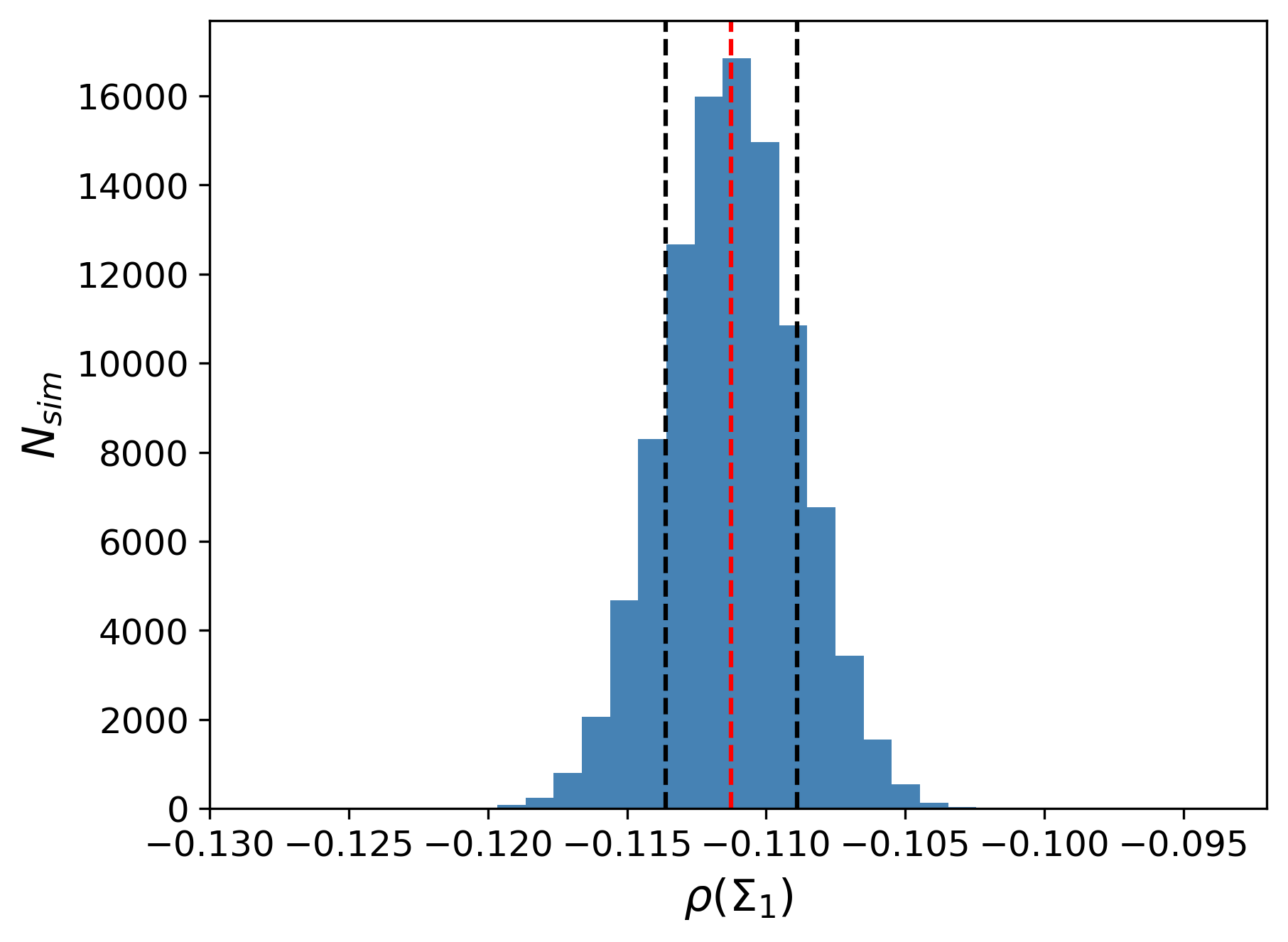}
\caption{The binned scatter plot between the gas-phase metallicity gradients and the central stellar mass surface densities. The number distribution of the correlation coefficient after bootstrapping is shown in the right panel. Marks and labels are the same as Figure \ref{fig:re}. \label{fig:s1}}
\end{figure*}

Subsequently, using the H-band S\'ersic index, stellar mass and effective radius, we calculate $\Sigma_1$, the projected stellar mass surface densities within 1 kpc radius ($\Sigma_1$). As plotted in Figure \ref{fig:s1}, there is no strong correlation between the central mass densities and the metallicity gradients, which is consistent with the lack of trend probed by S\'ersic indexes. The bootstrap distribution is narrowly centered with a 68\% confidence interval of [-0.114, -0.109].

\begin{figure}
\plotone{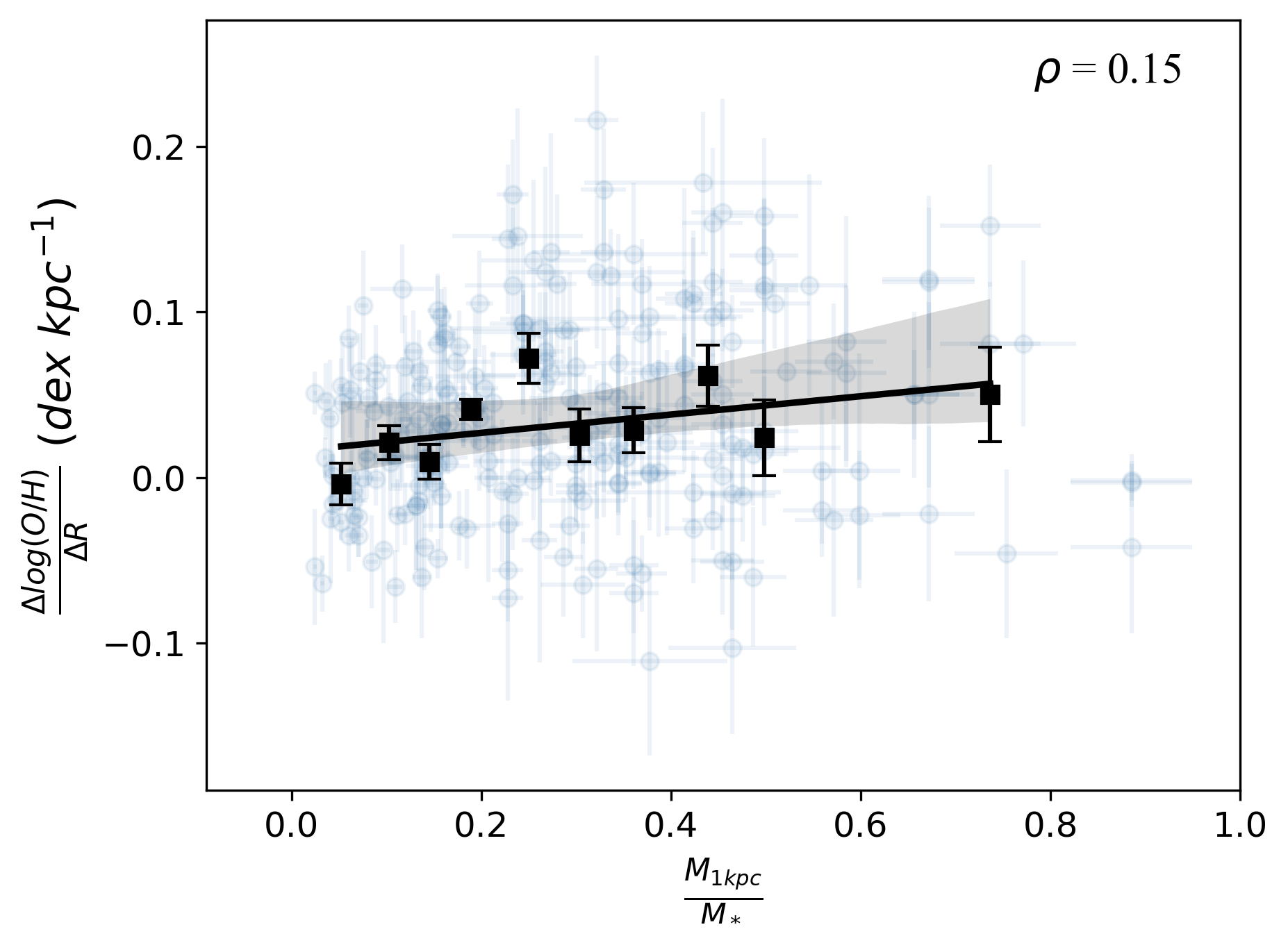}
\caption{The binned scatter plot between the gas-phase metallicity gradients and the normalized central stellar mass ($\frac{M_{1kpc}}{M_*}$). Marks and labels are the same as Figure \ref{fig:te}. \label{fig:s1_norm}}
\end{figure}

Given that $\Sigma_1$ strongly depends on the stellar mass, with more massive galaxies having larger central density, we also compare the metallicity gradients to the fraction of stellar mass within the central 1 kpc, $\frac{M_{1kpc}}{M_*}$, whose dependence on stellar mass is much weaker than that of $\Sigma_1$ \citep{2022ApJ...925...74J}. The relationship between the metallicity gradients and $\frac{M_{1kpc}}{M_*}$ is plotted in Figure \ref{fig:s1_norm}, which similarly shows no strong correlation.

With a significant correlation with stellar mass and tentative correlation with size, we detect null correlations with different measurements of concentration. This can be attributed to the large scatter in $R_e$ and $\Sigma1$ at a fixed stellar mass for star-forming galaxies. Promising candidates for the origin of this scatter are halo concentration and/or halo formation time \citep{2020ApJ...897..102C}.

\begin{figure*}
\plottwo{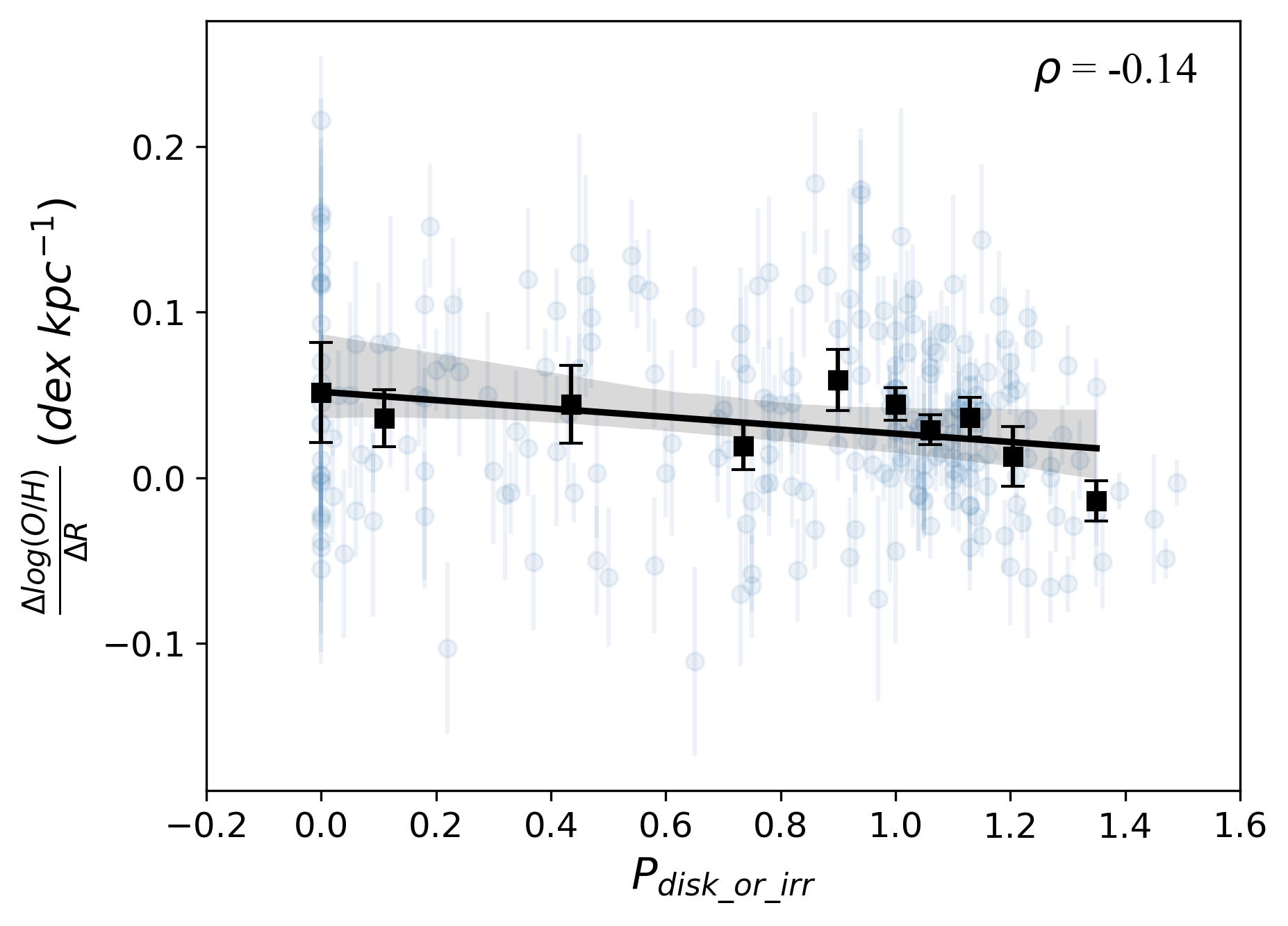}{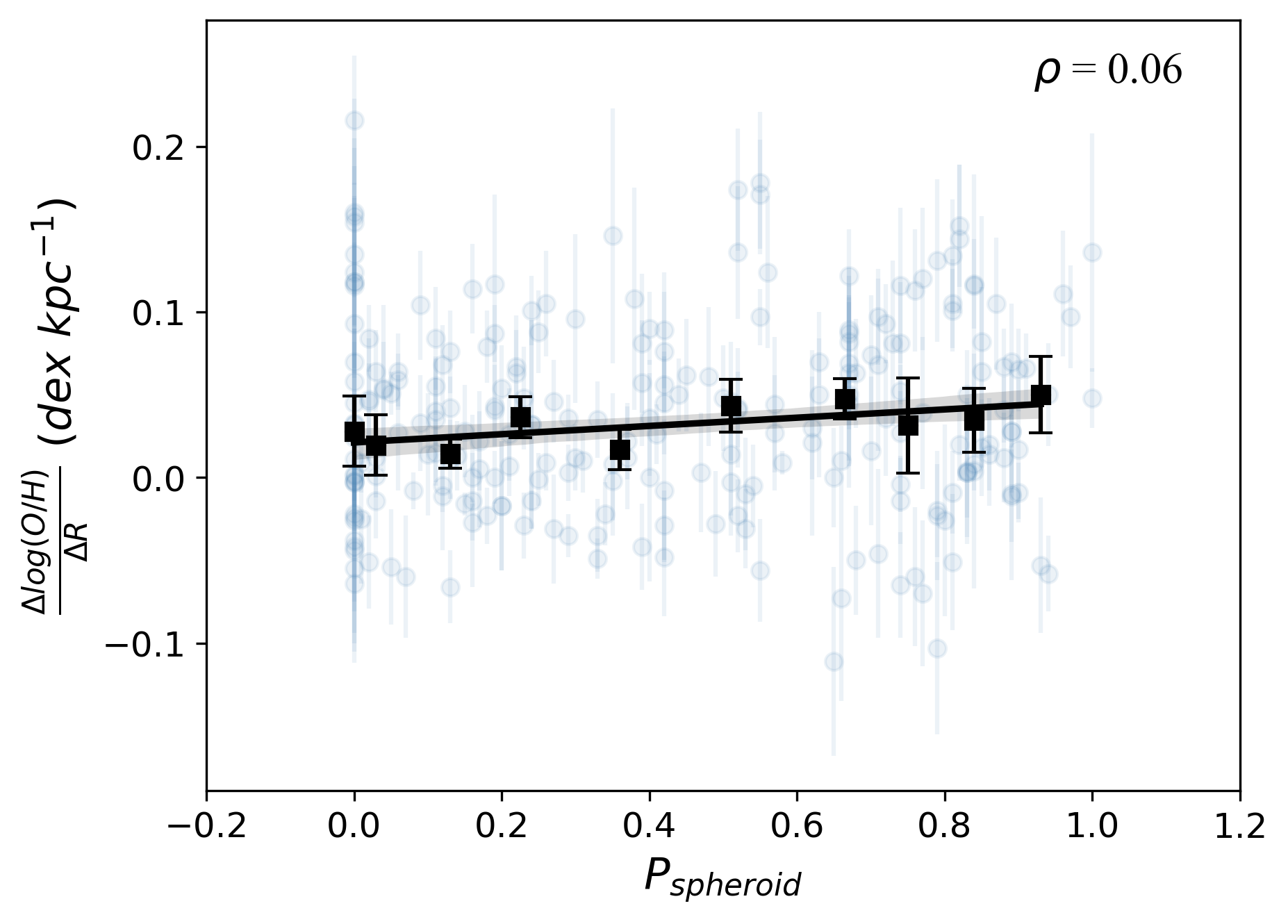}
\caption{The binned scatter plot between the gas-phase metallicity gradients and the morphologically classification probabilities. The left panel refers to the probability of being disk-like or irregular galaxies, the right panel refers to the probability of being spheroid galaxies. Marks and labels are the same as Figure \ref{fig:te}. \label{fig:mo}}
\vspace{1cm}
\end{figure*}

\begin{figure}
\plotone{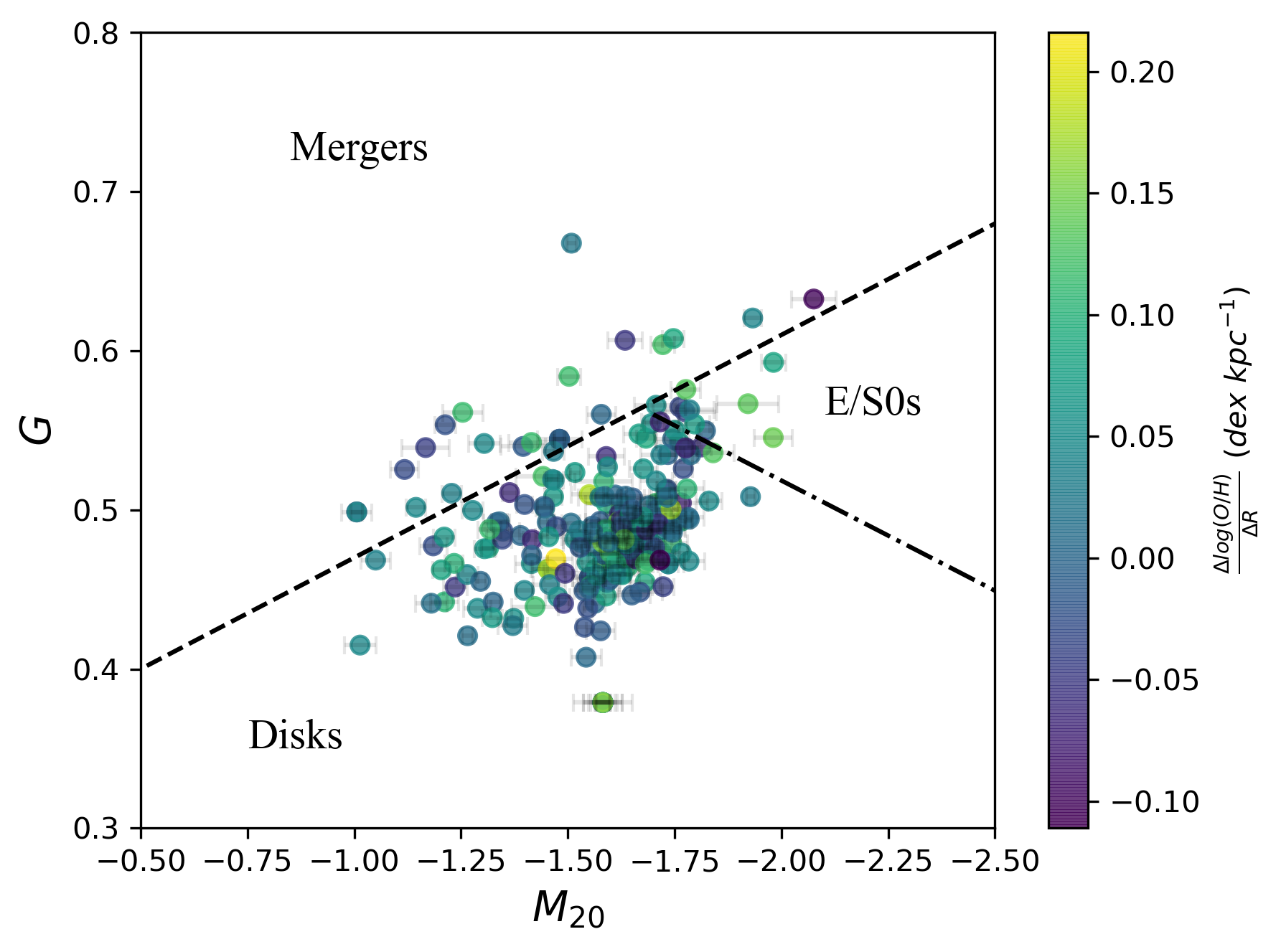}
\caption{The H band $Gini$ value as a function of $M_{20}$ color-coded by the gas-phase metallicity gradients. The black lines show the boundaries between mergers (dashed), disks and elliptical galaxies (dash-dotted) given by \citealt{2008ApJ...672..177L}. \label{fig:gini}}
\end{figure}

We have also used the visual classification of H-band morphologies given by \citealt{2015ApJS..221....8H}, which provides the probability of being spheroid/disk-like/irregular for each galaxy. Here we combine the disk and irregular probabilities into one single category for easier and more reliable comparison with spheroid probabilities. From Figure \ref{fig:mo}, the dependence of metallicity gradients on the visual morphology is rather weak.

The plane defined by the Gini coefficient $G$ and the second moment $M_{20}$, ($Gini$, $M_{20}$), provides a widely-used non-parametric classification of galaxy morphology, both regarding morphology as well as indication on the probability that they are mergers observed during various stages of the merging event (see, e.g.,  \citealt{2004AJ....128..163L,2008ApJ...672..177L}). Figure \ref{fig:gini} shows the distribution of our galaxies on the ($Gini$, $M_{20}$) plane. The lines mark the boundaries between mergers, disk-like galaxies and ellipticals/S0s, based on nearby galaxies and adjusted to z$\sim$1 relative to the calibration of the coefficient made at $z\approx 0$ \citep{2004AJ....128..163L,2008ApJ...672..177L}. It has been found, however, that there is no significant change in this diagnostic at  higher redshifts (e.g. \citealt{2022arXiv221014713K}). As seen in the plot, most of the galaxies in our sample have disk-like morphologies and relatively few of them are located in the region of the plane where candidate mergers are found. The figure also shows that there is a large dispersion  of metallicity gradients for both mergers and non-mergers (disk galaxies), which indicates that merging activities are not the dominant mechanism for the redistribution of gas and metals and the flattening of the metallicity gradients.

\section{Discussion} \label{sec:dis}

As shown in \citet{2021ApJ...923..203S}, the gas-phase metallicity gradients in star-forming galaxies of approximately the Milky Way mass at $0.6<z<2.6$ are essentially flat, with a broad scatter observed across the population (see Figure \ref{fig:hist}). This scatter accounts for galaxies with both negative and positive gradients, with the former being a minority of cases. The flat and mildly positive metallicity gradients of star-forming galaxies at the cosmic noon are in stark contrast with the measurements in the local universe, which in general show negative gradients \citep[e.g,][]{2019ApJ...887...80K, 2020ARA&A..58...99S}.

\begin{table*}[ht!]
\vspace{1cm}
\centering
\begin{tabular}{lllllllllll}
\hline\hline
\textbf{Parameter} & $t_{mw}$ & rSFR  & $\tau_{tot}$         & $\tau_1$       & $\tau_2$ & Skewness & $\frac{N_{multi}}{N_{tot}}$ & $t_E$         & $M_*$             & Color Gradient \\
\textbf{Coefficient} & 0.13  & -0.07 & 0.14             & 0.14         & 0.15   & -0.09    & -0.80           & -0.60        & -0.32            & -0.18          \\
\textbf{P-value} & 0.16 & 0.46 & 0.12 & 0.14 & 0.11 & 0.33 & 0.10 & 0.01 & $<$0.01 & 0.64 \\
\hline
\textbf{Parameter} & U-V   & V-J   & $A_V$(Prospector) & $A_V$(3D-HST) & n      & Re       & $\Sigma_1$        & $\frac{M_{1kpc}}{M_*}$ & $P_{disk\_or\_irr}$ & $P_{spheroid}$    \\
\textbf{Coefficient} & -0.05 & -0.17 & -0.09            & -0.28        & -0.01  & -0.21    & -0.11           & 0.15         & -0.14            & 0.06   \\
\textbf{P-value} & 0.47 & 0.01 & 0.33 & $<$0.01 & 0.62 & $<$0.01 & 0.05 & $<$0.01 & 0.02 & 0.24 \\
\hline\hline
\end{tabular}
\caption{The Spearman’s rank correlation coefficients and p-values between the gas-phase metallicity gradients and different galaxy physical properties.} \label{tab:sum}
\end{table*}

Overall, we find that the gas-phase metallicity gradients show rather weak correlations with the properties of galaxies, which is in agreement with the analysis by \citealt{2021ApJ...923..203S}. A summary of all the Spearman’s rank correlation coefficients and p-values is given in Table \ref{tab:sum}. From new measures of the stellar mass (total mass formed) that we have obtained from SED modeling with $Prospector$, we find a noisy but significant anti-correlation between the metallicity gradient and the stellar mass. Namely, more massive star-forming galaxies have less positive (flatter and negative)  metallicity gradients, although the scatter is large at lower stellar mass. The dependence of metallicity gradients on the stellar mass has been widely discussed  (e.g., \citealt{2016MNRAS.463.2513B, 2019ApJ...872..144H}). Low-mass galaxies and active star-forming galaxies tend to have flat gradients, likely due to more efficient stellar feedback \citep{2017MNRAS.466.4780M}. \citealt{2016MNRAS.463.2513B} argue that the resolved mass-metallicity relation can reproduce the gas-phase metallicity gradients for all galaxies except the least massive ones. The stellar-mass range of galaxies in our sample is relatively narrow (8.5 $<~log~M_*/M_{\odot}~<$ 10.5) and falls in the low-mass end of the present-day mass function, which may explain the general flatness and lack of correlations in the gradients, given the relatively small dynamic range. 

Given that star-forming galaxies increase their stellar mass as they evolve, another way to interpret this anti-correlation is that as the stellar mass of galaxies grows with time, the gas-phase metallicity gradients decrease. 
Additional evidence in support of this trend is that galaxies with more negative gas-phase metallicity gradients are more likely to have experienced multiple star forming episodes in their SFH, which also requires longer evolutionary time. 

Further independent evidence comes from the weak anti-correlation between the metallicity gradient and the size of galaxy as quantified by $R_e$, which increases with time in star-forming galaxies as they evolve (e.g. \citealt{2017ApJ...834L..11A, 2019A&A...625A.114J}). Note that the correlation with $R_e$ is most likely diluted by the large scatter in $R_e$ at fixed stellar mass for star-forming galaxies at high redshifts \citep{2020ApJ...897..102C}. The growth of $R_e$ can be directly observed in the left panel of Figure \ref{fig:te}, where $R_e$ is plotted against the evolutionary time $t_E$, providing a metric of the amount of the evolution a given galaxy has undergone through at the epoch of observation ($z_{obs}$) of observation (Giavalisco et al. in prep; see also the `mass doubling number' in \citealt{2022ApJ...926..134T}). 

Lastly, additional evidence that the evolutionary trends of metallicity gradients  comes from the anti-correlation between the metallicity gradient and the evolutionary time $t_E$ itself, which is shown in the right panel of Figure \ref{fig:te}. This anti-correlation directly reflects the `flattening' or the decrease of the gas-phase metallicity gradients with time. In conclusion, the implications from all these anti-correlations are in good agreement with each other, indicating that star-forming galaxies develop flatter or more negative metallicity gradients as they evolve.

Except for the correlation illustrated above, we find no additional statistically-significant correlation between the gas-phase metallicity gradients and other stellar population properties of the galaxies, such as the recent SFR, mass-weighted age, dust attenuation, rest-frame UV and optical colors. This finding aligns with earlier results reported in \citet{2021ApJ...923..203S}, where no strong correlations are found between the metallicity gradients and SFR, sizes, star-formation surface density, and star-formation per potential energy. Clearly, the generally negative gradients observed in the local universe must be the result of evolutionary process that took place after the epochs of observation reported here, namely $0.6<z<2.6$. The lack of correlation with rest-frame colors and color gradients described here also suggests that the stellar age or stellar activities in the broad picture can not fully explain the observed metallicity gradients. Dust attenuation plays a marginally important role to shape the metallicity gradients, where more dusty galaxies grow or retain more negative metallicity gradients, but the correlation is quite noisy, especially at the low dust attenuation end. 


As mentioned in Section \ref{sec:intro}, due to the local enrichment from star-formation, a negative radial gradient in the gas-phase metallicity would be a natural consequence of inside-out galaxy formation scenarios. However, gas exchanges continuously take place in galaxies on rapid timescales (e.g. \citealt{2015MNRAS.453.4337S}), including outflows of enriched gas and inflows of gas at low metallicity  accreted from outside the virial radius. These gaseous exchanges evidently contribute to shaping the evolution of the metallicity gradients. The magnitude of such effects, of course, depends on the extent of radial gaseous flows. The phenomenology presented in this paper suggest that star-forming galaxies evolve from an early evolutionary phase characterized by flat or mildly positive gradients toward more evolved phases during which the metallicity gradients are essentially flat or negative. This dependence is not necessarily in contrast with the inside-out galaxy formation scenario, it just evidences the effects of large-scale redistribution of metals in the early stages of galaxy evolution.

The broad dispersion of metallicity gradients at high redshifts, in contrast with observations in the local universe, adds to the complexity of understanding how the metallicity gradients are affected by the assembly history of galaxies. Given the generally higher and sustained star formation rates, the big diversity of gas-phase metallicity gradients observed at high-z would be naturally explained by the negative gradients being rapidly destroyed before fully established, so that the well defined negative gradients are rare at among the general population of star-forming galaxies. 

Conversely, the flat and mildly positive metallicity gradients may trace accretion or radial inflow of near-pristine gas. Accretion of metal-depleted gas and efficient transport of this gas toward the center of the galaxies, which is intimately related to metal transports, can also efficiently flatten or even reverse gas-phase metallicity gradients, mainly through winds, accretion, stellar feedback, and merging events \citep{2017MNRAS.466.4780M, 2019MNRAS.490.4786G, 2020MNRAS.498.1939G}. Note that our analysis is based on a sample of active star-forming galaxies which mostly include disks. Recently, \citet{2022arXiv221002478B} report that early-type galaxies with lower rotational support tend to have more negative metallicity gradients. 

As the galaxies evolve, however, our analysis suggests that the rate of change of metallicity due to star-formation is higher in the centers of galaxies than in their outskirts (see the toy model in \citealt{2021ApJ...923..203S}), and thus negative metallicity gradients would naturally develop. Losing metal-rich gas preferentially from the outskirt, where the escape velocity is lower, and accreting high-angular-momentum gas would also help decrease the gas-phase metallicity gradients. \citet{2023arXiv230502959G} find significant evidence of non-tangential flows in star-forming galaxies at z$\sim$2, where the gravitational torques drive gas rapidly inward and result in the formation of central disks and large bulges. Similarly, \citet{2023arXiv230516394W} report the discovery of cool gas inflows towards three star-forming galaxies at z$\sim$2.3, their analysis suggests that the inflows are due to recycling metal-enriched gas from previous ejections.

We now attempt to develop some physical explanations for the lack of correlations with some of stellar population properties discussed above. First of all, the measured metallicity gradients in our sample don't show big deviation from flat gradients. The overall range is [-0.10, 0.16] dex kpc$^{-1}$, with more than 60\% galaxies have flat gradients (within [-0.05, 0.05] dex kpc$^{-1}$). Considering the uncertainties of measurements, the small diversity of gradient values disfavors strong correlations with any galaxy properties.

Some works claim that the total stellar mass plus the local stellar mass surface density are sufficient to qualitatively reproduce the observed trends of metallicity gradients (e.g. \citealt{2021MNRAS.501..948B}). However, whether we can use similar relations to understand the gas-phase metallicity gradients in high-z galaxies is questionable. At early times, it is suggested by the flat metallicity gradients that metals are more frequently and efficiently redistributed due to the prevalence of gas transport mechanisms such as gas accretion, outflows and mergers. In parallel, the elevated levels of gas turbulence detected in the interstellar gas of galaxies at $z\,>0.5$ \citep{weiner06, FS06, FS09, kassin07, kassin12, wisnioski15, simons16, simons17, ubler19, price20} will promote elevated mixing of the gas-phase metals in galaxies.

The kinematic properties of high-z galaxies can vary on a relatively fast time scale together with their star-formation activities and feedback, and then significantly change the gas-phase metallicity gradient. 
Note that \citet{2023arXiv230502959G} reports evidence of radial gaseous flows based on the deviations of the gas kinematics from differential rotation.

Thus, the general lack of strong correlations between metallicity gradients and a number of properties related to the star-formation activity presented here suggests that the observed metallicity gradients could reflect more the instantaneous star formation distribution and rapid gaseous flows rather  than the long-term growth history of high-z galaxies. This conclusion also finds support by numerical simulations (e.g., \citealt{2017MNRAS.466.4780M}).

\section{Summary} \label{sec:sum}

In this paper, we use the Bayesian SED fitting code $Prospector$ to reconstruct the non-parametric SFHs of 238 star-forming galaxies, and explore possible relations between a number of galaxy properties and the gas-phase metallicity gradients provided by the CLEAR survey \citep{2021ApJ...923..203S}. Our main conclusions are summarized below:

\begin{itemize}

\item In our sample, there is a weak but significant anti-correlation between the metallicity gradient and the stellar mass (see Figure \ref{fig:mass_mul}), suggesting that the gas-phase metallicity gradients decrease as the stellar mass of galaxies grows with time. 

\item There is an anti-correlation between the metallicity gradient and the evolutionary time (Figure \ref{fig:te}), suggesting that less evolved galaxies tend to have more positive metallicity gradients, these gradients become progressively smaller and possibly negative as the galaxies evolve.

\item Galaxies with smaller metallicity gradients are more likely to have experienced multiple star forming episodes in their SFH (Figure \ref{fig:mul}).

\item We find no statistically-significant correlation between the gas-phase metallicity gradients and other stellar population properties of the galaxies ($|$Spearman's rank correlation coefficients$|<0.2$), including the mass-weighted age, recent star formation rate, the overall SFH shape (e.g., total time width, asymmetry), dust attenuation, rest-frame UV/optical colors, and the morphology (see analysis in Section  \ref{sec:sfh}, \ref{sec:color}, and \ref{sec:morph}). The lack of correlations could be partly due to the narrow stellar mass range, which also extends toward the low-mass end. 

\end{itemize}

While the metallicity gradients of nearby galaxies can be understood by certain relations, the situation is much more complex in the high-z universe. At early time, gas kinematics can frequently change the distribution of metals in and around galaxies with a relatively short timescale, which means that the observed metallicity gradients may only reflect the instantaneous state of galaxies at the time of observation (see e.g., Acharyya et al. in prep).

These observed correlations or, better, lack of correlations, can help constrain theoretical models of galaxy evolution and be compared with numerical simulations. 
The GOODS-N and GOODS-S fields have been probed with several JWST surveys, including the JWST Advanced Deep Extragalactic Survey (JADES; \citet{2023arXiv230602465E}), the JWST Extragalactic Medium-band Survey (JEMS; \citet{2023ApJS..268...64W}), and the First Reionization Epoch Spectroscopic COmplete Survey (FRESCO; \citet{2023MNRAS.525.2864O}). In the near future, by incorporating the JWST photometry, we will be able to revisit our analysis with larger data sample and more reliable SED modeling. Given high-quality spatial resolved metallicity measurements at high redshifts, robust comparison between the data and models can be realised.

\section{Acknowledgement}
We thank our collaborators from the CLEAR team for many valuable discussions and contributions.
This work has made use of the Rainbow Cosmological Surveys Database, which is operated by the Universidad Complutense de Madrid (UCM), partnered with the University of California Observatories at Santa Cruz (UCO/Lick,UCSC).

The parallel SED fittings are performed by the Unity cluster, which is a collaborative, multi-institutional high-performance computing cluster located at the Massachusetts Green High Performance Computing Center (MGHPCC).
\software{Astropy \citep{2013A&A...558A..33A,2018AJ....156..123A,2022ApJ...935..167A},  
          FSPS \citep{2009ApJ...699..486C,2010ApJ...712..833C},
          dynesty \citep{2020MNRAS.493.3132S},
          Prospector \citep{2021ApJS..254...22J}, 
          }

\appendix

\section{Spectral Line Detection and Fitting}
\label{appendix:a}

\begin{figure}
\plotone{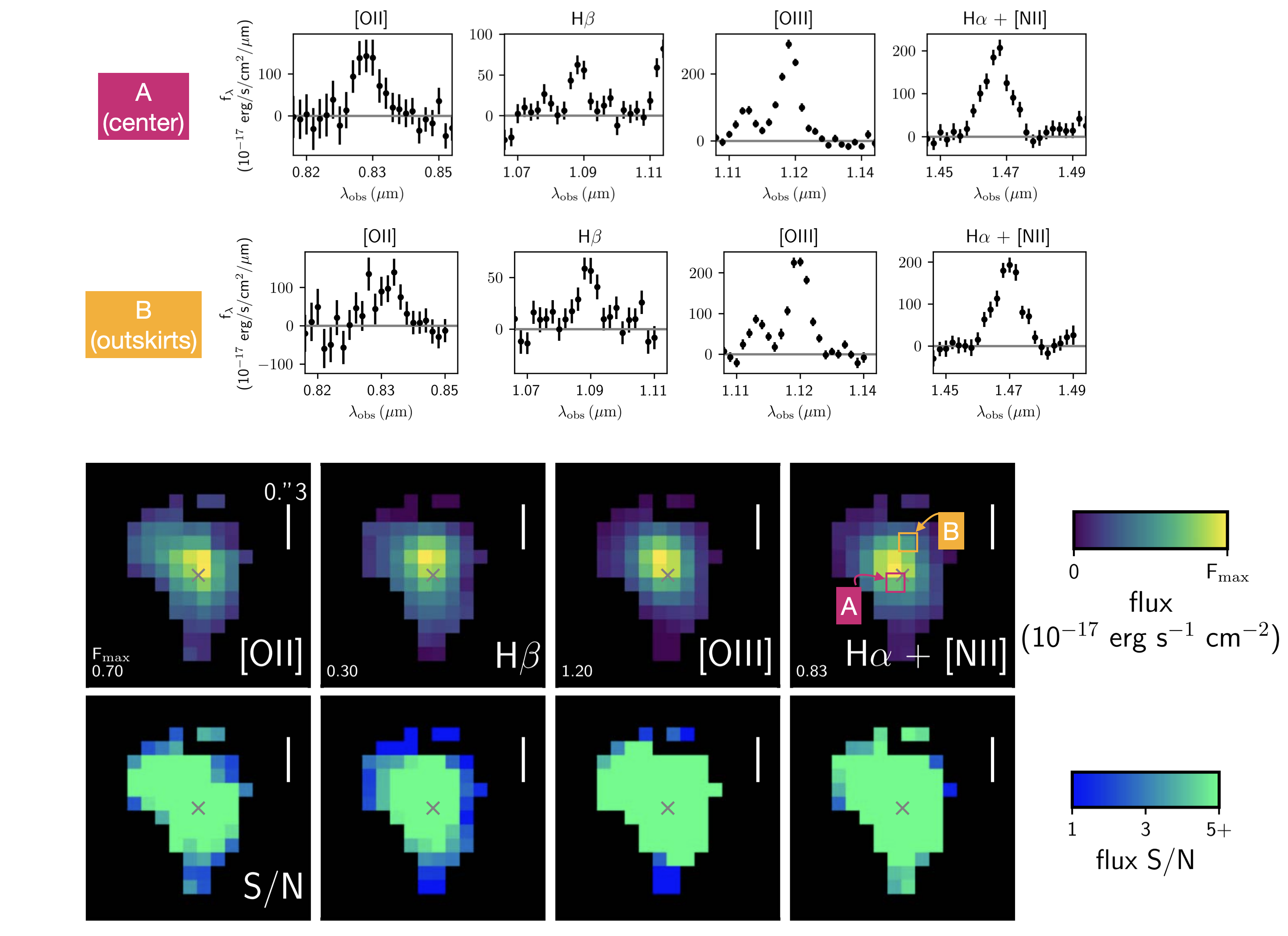}
\caption{Examples of line maps and spatially-resolved 1D extracted spectra from the CLEAR observations. The data have been presented and discussed by \citet{2021ApJ...923..203S}, who also provide a complete description of the data reduction, analysis and spectral extraction procedures. In this paper we have used exactly the same data set. The lower panel shows the 2D spectra and S/N maps for one of the sample galaxies. The magenta and orange boxes mark examples of the extraction regions for the center and outskirt respectively. The gradients have been obtained after azimuthal averaging. The extracted 1D spectra are plotted in the upper panel, where the spectral features are clearly identified. \label{fig:2dspec}}
\end{figure}

The data used in this paper are presented and described in the paper by \citet{2021ApJ...923..203S}, including data reduction, analysis, extraction of the spectra and measures of the metallicity gradients. Here we provide some examples of the emission line maps and extracted spectra to illustrate the data. In Figure \ref{fig:2dspec}, we show 2D maps of the optical lines of one galaxy from our sample, color coded by pixel SNR, as well as the 1D extracted spectra of the lines in selected extraction boxes. \citet{2021ApJ...923..203S} also detail the fitting procedures to measure the line properties, as well as the azimuthal averaging algorithm to derive the metallicity gradients.

\section{Robustness Test of PROSPECTOR Fittings}
\label{appendix:b}
Different choices of the input photometry bands and the sampling of the lookback time could affect the SED fitting results, and in turn the reconstructed SFHs. To test the stability and robustness of our $Prospector$ fittings, we compare the output SFHs with different numbers of input photometry bands and lookback time bins.

\begin{figure}
\plottwo{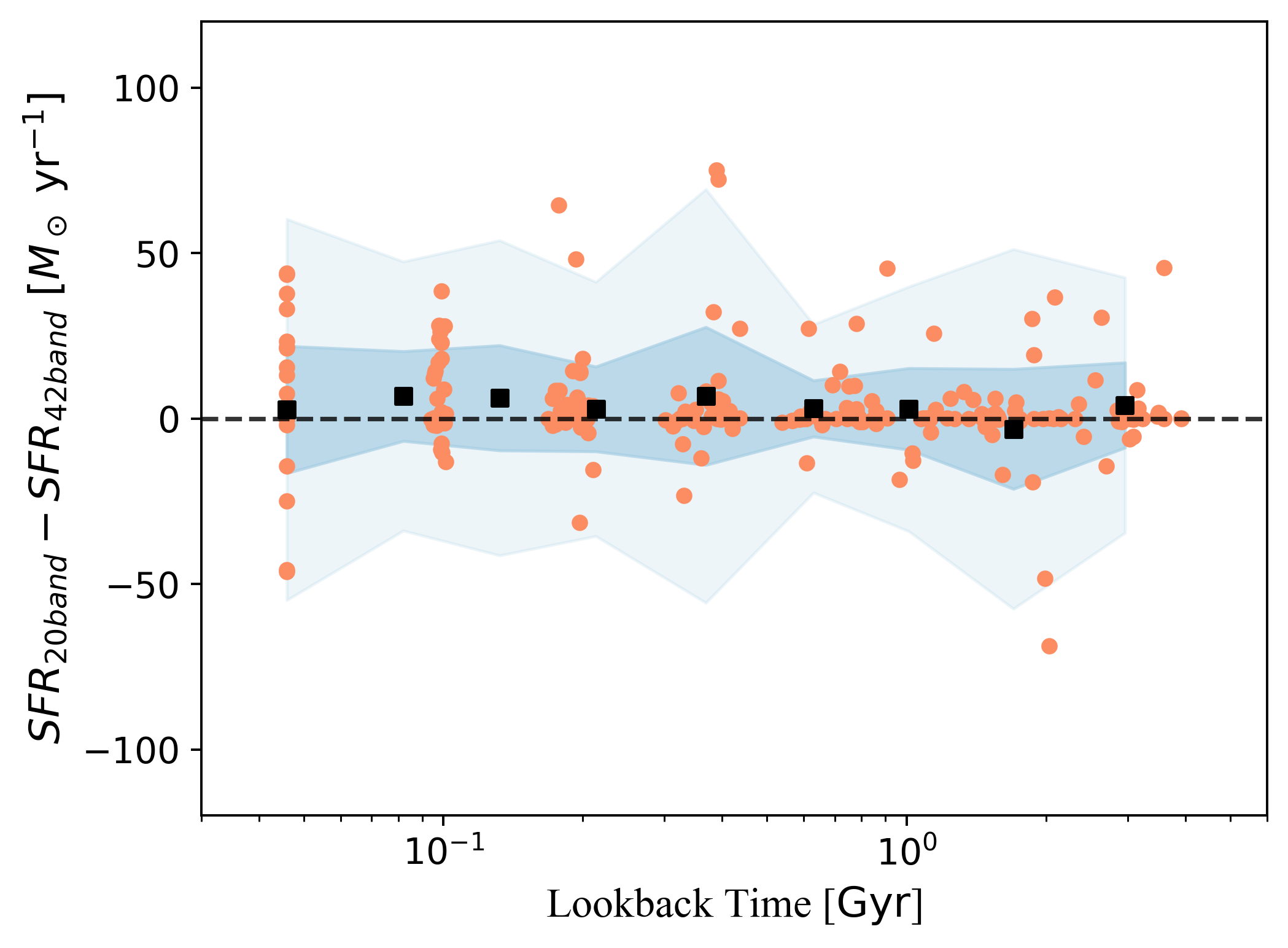}{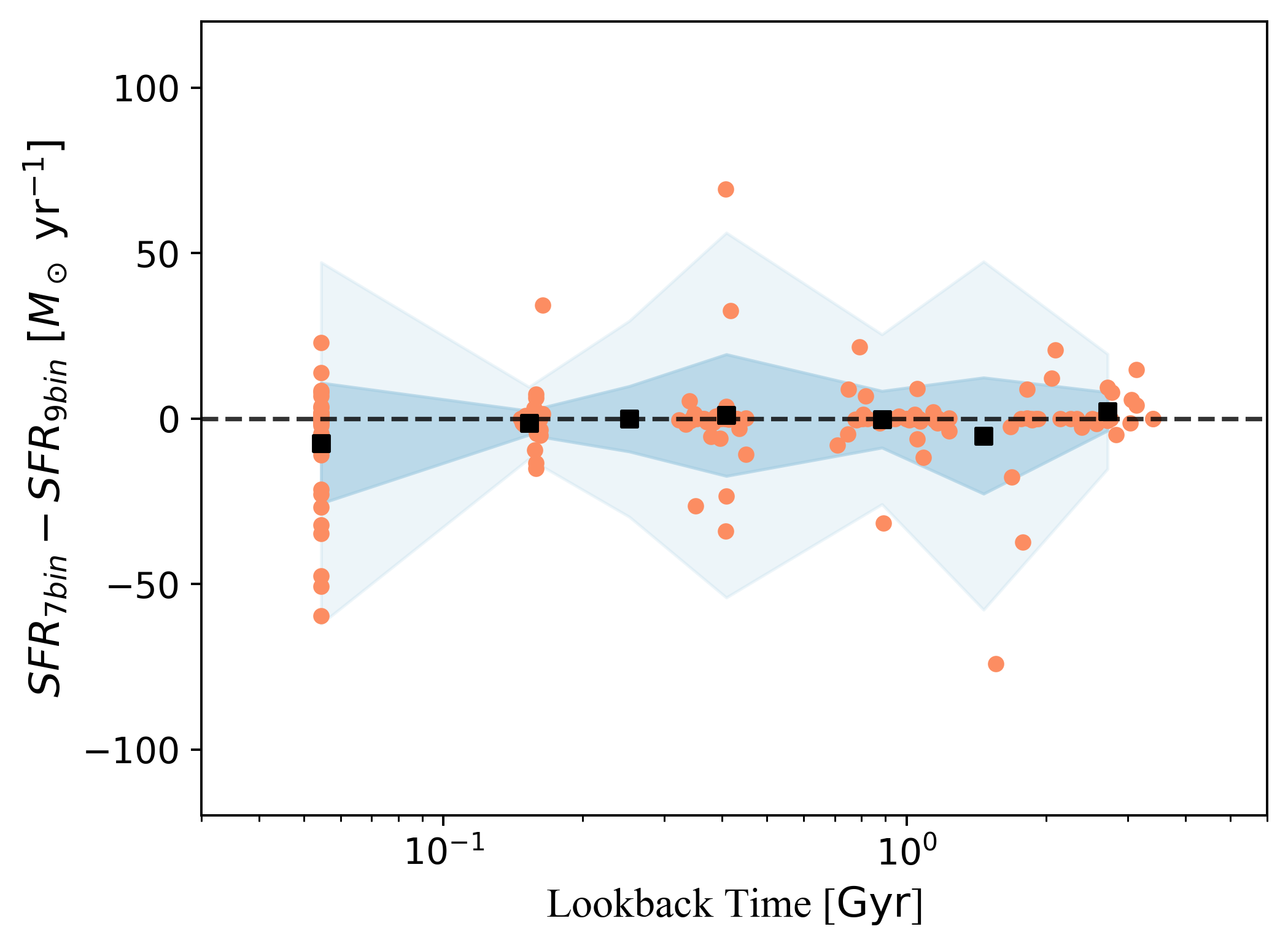}
\caption{The differences between the reconstructed SFHs in terms of different photometry band choices (left) and lookback time bin choices (right). The orange scattered points represent individual galaxies, the black squares indicate the median Y axis values for each time bin, and the shaded areas show the 68th and 95th percentiles. In the right panel, to directly compare the 9-bin results with the 7-bin results, the SFR values have been interpolated from adjacent bins.  \label{fig:diff}}
\end{figure}

As described in Section \ref{sec:fit}, we adopt the $\sim$20 photometry bands provided by the 3DHST survey to test the SFH dependence on the input photometry bands. The results are highly consistent for most galaxies in our sample, while a few of them show different features in the best-fit SFH. Particularly, comparing with the original choice of $\sim$40 photometry bands, there is less variation in the reconstructed SFH among different galaxies when using $\sim$20 bands. As shown in Figure \ref{fig:diff}, there is no systematic trend in the SFH difference between the two. We conclude that the optimization could be slightly undermined by local minima when fewer photometry bands are available. Therefore, we decide on using all available photometry bands and filter those sources with scarce data points.

For the test of time bins, aside from the original setting of 9 bins, we adopt the 7 time bins defined by \citealt{2019ApJ...876....3L} and adjust them based on the age of the universe at the known redshift. To compare the 7-bin SFH of the test runs with the 9-bin SFH from our primary runs, we perform a linear interpolation of the 9-bin SFHs to make equivalent 7-bin results. The differences in the reconstructed SFHs are shown in the right panel of Figure \ref{fig:diff}. The results are highly consistent for different choices of time resolution and there is no systematic trend in the SFH differences.

In summary, the test runs indicate that the reconstruction of SFHs from $Prospector$ is stable against the input photometry and the time resolution, supporting the robustness of our results.

\begin{figure}
\includegraphics[scale=0.46]{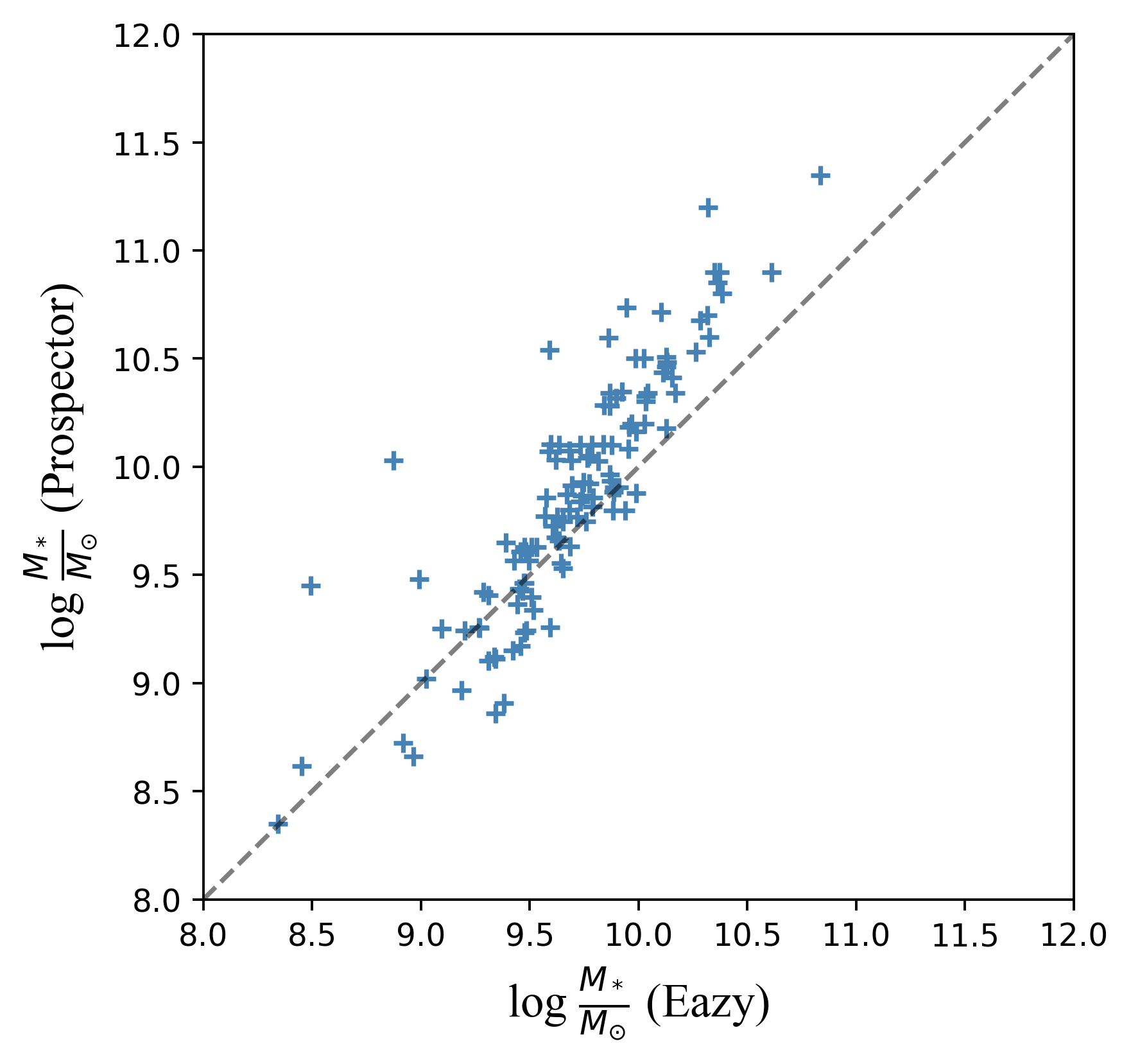}
\includegraphics[scale=0.47]{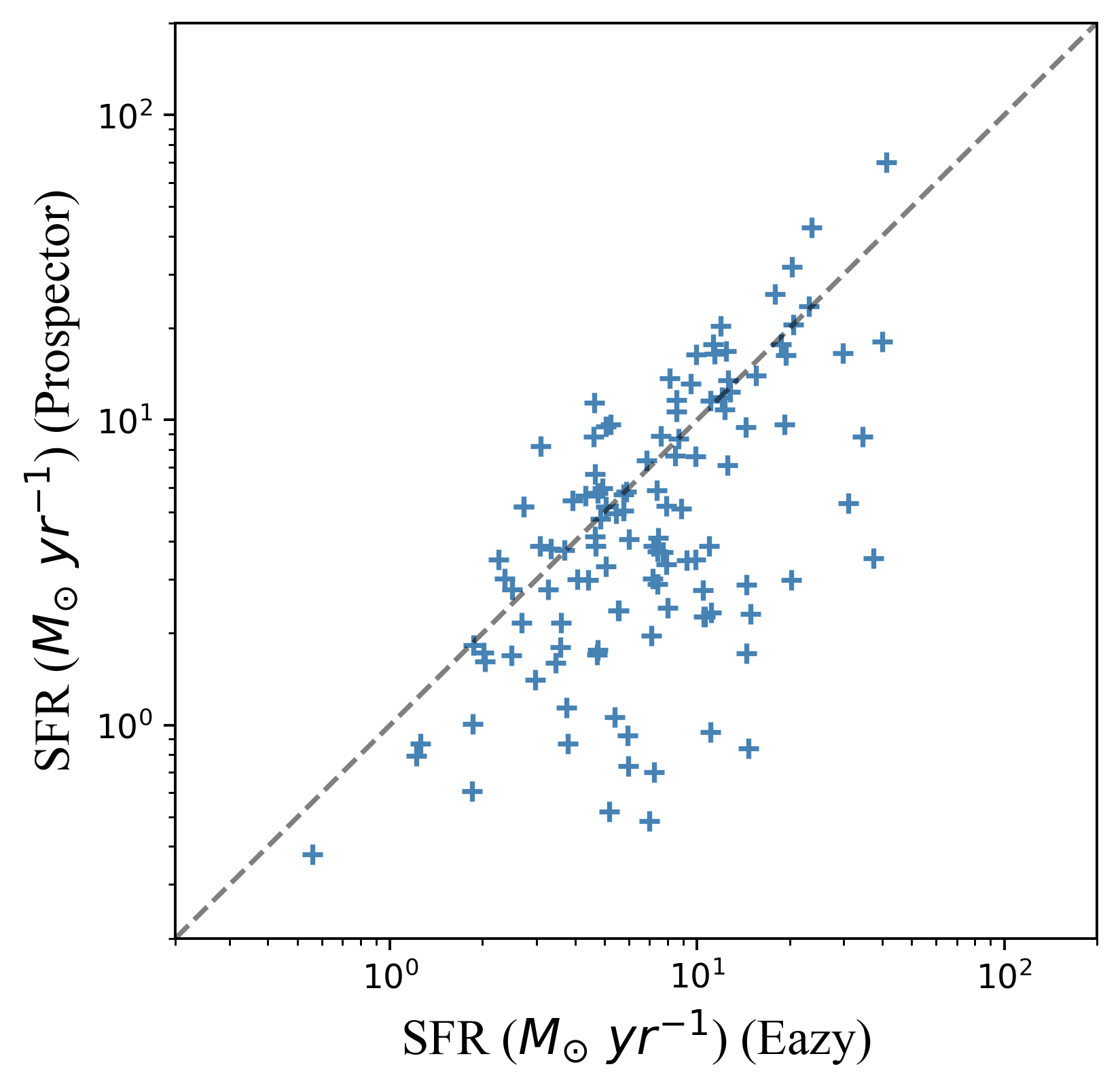}
\includegraphics[scale=0.47]{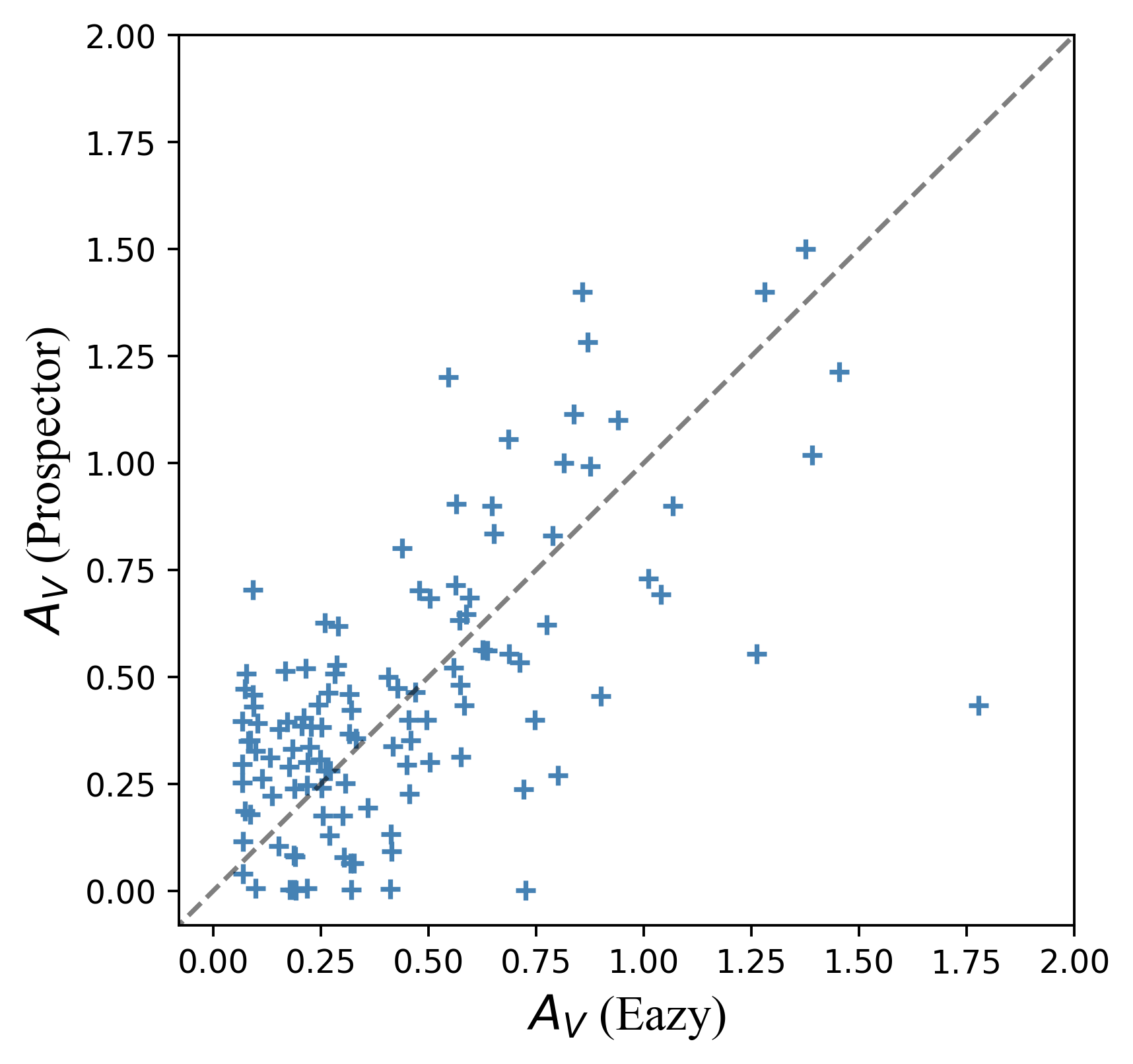}
\caption{Comparison of SED fitting outputs between EAZY (\citet{2021ApJ...923..203S}) and $Prospector$ (this work). The dashed line marks the one-to-one relation. \label{fig:eazy}}
\end{figure}

Furthermore, to study the differences between SED fitting codes, we compare our fitting results with those reported by \citet{2021ApJ...923..203S}. Their SED modeling is conducted by EAZY \citep{2008ApJ...686.1503B}, which is optimized for producing high-quality photometric redshifts over $0<z<4$. The comparison of output stellar masses, star formation rates, as well as the dust attenuation $A_V$ is shown in Figure \ref{fig:eazy}. $Prospector$ tends to give higher stellar mass than EAZY at the high mass end, and provides relatively lower SFR, while the dust attenuation is uniformly scattered.
The differences between multiple SED fitting codes have long been identified and discussed in literature (e.g., \citealt{2019ApJ...877..140L, 2023ApJ...944..141P}). Here we find no significative differences bias in the results of SED modeling from the two codes.

\bibliography{draft}{}
\bibliographystyle{aasjournal}

\end{document}